# Silica-coated admixtures of bismuth and gadolinium oxides for 3D printed concrete applications: rheology, hydration, strength, microstructure, and radiation shielding perspective

Pawel Sikora[1,*], Szymon Skibicki[1], Mehdi Chougan[2], Piotr Szewczyk[1], Krzysztof Cendrowski[1], Karol Federowicz[1], Ahmed M. El-Khayatt[3], H. A. Saudi[4], Jarosław Strzałkowski[1], Mohamed Abd Elrahman[5], Mateusz Techman[1], Daniel Sibera[1], Sang-Yeop Chung[6]

[1] Faculty of Civil and Environmental Engineering, West Pomeranian University of Technology in Szczecin, Poland
[2] Department of Civil and Structural Engineering, The University of Sheffield, Sheffield S1 3JD, United Kingdom
[3] Department of Physics, College of Science, Imam Mohammad Ibn Saud Islamic University, (IMSIU), Riyadh, Saudi Arabia
[4] Department of Physics, Faculty of Science, Al-Azhar University, Women Branch, Nasr City, Cairo, Egypt
[5] Structural Engineering Department, Mansoura University, Mansoura City 35516, Egypt
[6] Department of Civil and Environmental Engineering, Yonsei University, Seoul 03722, Republic of Korea

**ABSTRACT**

This study examines the impact of replacing up to 5 vol% of Portland cement (PC) with both pristine $Bi_2O_3/Gd_2O_3$ (BG) and silica-coated BG particles. Two different types of silica coatings, each with varying synthesis methods, were applied to coat the BG structures, and their impact on the fresh, hardened, microstructure and radiation-shielding performances of the 3D printable cementitious composites (3DPC) was investigated. Isothermal calorimetry demonstrated that pristine BG incorporation delays hydration, whereas silica coatings mitigate this, with type A coating being more effective. Early compressive strength was reduced in BG-containing mixes but normalised after seven days. Rheological tests showed that BG additives enhanced thixotropy and yield shear stresses, with 2.5 vol% being optimal, especially with method B coating. Green strength properties improved significantly with method B coated particles, showing up to 62.4% and 57.7% increases in strength and modulus, respectively, after 30 minutes. Micro-CT and MIP analyses confirmed reduced porosity and refined pore structure with silica coatings. Radiation shielding tests indicated superior performance in uncoated BG mixes, with method B coating providing superior shielding performance compared to that of method A coatings due to their higher surface area. In general, silica-coated BG particles enhance the mechanical, rheological, and radiation shielding properties of 3DPC, with method B coatings offering the most significant benefits.

**Corresponding authors:**
*Pawel Sikora – email: pawel.sikora@zut.edu.pl; Faculty of Civil and Environmental Engineering, West Pomeranian University of Technology in Szczecin, Szczecin, Poland.

# 1. Introduction

3D concrete printing (3DCP) is an innovative technology within Industry 4.0 and digital fabrication. Thanks to geometrical freedom, complex structures with sophisticated topologies that would be difficult or impossible to produce using traditional construction methods can be designed and printed. It is limited to entire structures and particular elements, e.g., bricks or building envelopes with optimized thermal/mechanical performances [1]. Moreover, 3D printing has attracted particular attention in producing protective structures and lunar applications [2–4]. To date, the 3D printing of polymers has received spectacular attention in space engineering and the production of various shielding elements for space and earth applications [5]. Particular emphasis has been placed on developing radiation safety materials by modifying filaments/inks with various attenuating components such as $B_4C$, $Gd_2O_3$ or bismuth particles [6–8]. Thus, it is now possible to produce complex topology structures with optimized ionizing radiation, electromagnetic shielding efficiency, or both [9].

Concrete is one of the most popular materials used to produce biological shields to protect humans from ionizing and electromagnetic radiation in hospitals, medical facilities, and nuclear power plants [10,11]. The improvement of attenuating performance can be achieved by proper introduction, most preferably heavyweight aggregates and additives. The use of 3DPC for radiation shielding has manifold benefits, including possibilities of optimizing the topology of 3D printed attenuating structures, remote access with robotic arms as well as potential operation in extreme conditions (elevated radiation or temperature); thus printing can be performed with minimal personal involvement. The substitution of aggregates in 3DPC is generally less appealing from a technological standpoint, as demonstrated by our prior research [12]. However, replacing them with fine powders as a binding agent poses a significant risk to the crucial properties of 3DPC, such as hydration and its early rheological performance necessary for the correct printing process. It is pivotal to use a combination of particles countering specific radiation sources to sufficiently attenuate both types of radiation (gamma-ray and neutron). In brief, to ensure adequate gamma-ray attenuation, materials with a high atomic number (Z) are required to facilitate higher interaction probability with photons, while materials with high neutron cross-sections are required to attenuate neutrons [13].

Therefore, multiple groups follow the principle previously introduced in polymer technology: introducing various combinations of micro-sized and nano-sized materials into a cementitious system. For instance, Soni et al. [14] combined WC and $B_4C$ particles in concrete, while Nikbin et al. [15] proposed the introduction of nano-$Bi_2O_3$ and nano-$WO_3$ particles for targeted improvement of radiation shielding. The incorporation of nano-sized particles leads to remarkable improvements in attenuating properties attributable to this behavior are (1) surface-to-volume ratio and (2) quantum confinement effects [16]. Similarly, various comparative studies on cement-based materials have already confirmed the superiority of nano-sized PbO [17], $WO_3$ [17], and $Bi_2O_3$ [18,19] over coarser (microsized) particles. Parallelly, researchers have paid high interest in developing advanced functional fillers such as W/$Gd_2O_3$ structures [20,21] or cobalt-doped titania nanocomposites [22]. Such an approach enables tailoring and optimizing the performance of the additive within the matrix on the molecular level, along with facilitating its introduction. The latter approach has not been deeply investigated in concrete technology. However, Ibrahim et al. [23] have recently successfully developed $CeO_2$/$ZrO_2$ nanocomposites to improve the gamma-ray attenuation properties of cement pastes.

This study introduces a novel approach to creating advanced functional admixtures using a combination of $Bi_2O_3$, $Gd_2O_3$, and $SiO_2$ for use in 3D printed concrete for radiation shielding. Combining these particles enhances the shielding capabilities for gamma-rays and neutrons while improving the rheological properties.

## 2. Research significance and scope

The knowledge of radiation shielding 3D printed concrete production is limited, and very few achievements have been made. Additionally, no knowledge is available on the effects of heavyweight admixtures on the hydration, fresh, and rheological properties of 3DPC. It is worth noting that the selected radiation shielding particles have detrimental effects on cementitious systems such as the highly popular $Bi_2O_3$, which is considered a suitable environmentally friendly and non-toxic lead alternative. However, $Bi_2O_3$ dramatically delays the hydration process of cement, thus leading to substantial retardation of setting and strength development [24,25]. Therefore, from the perspective of 3DPC, it is pivotal to produce admixtures adjusted to suitable performance during the deposition state of fresh 3DPC and in a hardened state.

This study aims to develop a functional admixture for 3DPC composed of $Bi_2O_3$ and $Gd_2O_3$ particles and $SiO_2$ coating for addressing both radiation shielding types (gamma ray – $Bi_2O_3$ and $Gd_2O_3$ – neutron) and ensuring satisfactory rheological and hydration performance in early hours of hydration ($SiO_2$). For this study, $Bi_2O_3$ was chosen, which is one of the most common gamma-ray attenuating admixtures along with $Gd_2O_3$, which is a material with extraordinary thermal neutrons absorbing capacity ($^{157}Gd$, with a natural abundance of 15.7% and possesses the highest thermal neutron capture cross-section among all known stable isotopes [26]). In order to produce such admixtures, a sol-gel silica coating synthesis method was proposed. The authors and other research groups have already explored this approach to produce silica-coated $Fe_3O_4$ [27], $TiO_2$ [28], carbon nanotubes [29], and steel fibers [30] for cement-based composite applications.

This study proposed two approaches to silica coatings: varied silica morphology, reactivity, and specific surface area. Afterward, synthesized functional admixtures were subjected to a comprehensive experimental campaign in 3D printed concrete material. The campaign has been divided into two main parts related to determining optimal admixture dosages: rheological, hydration, and mechanical evaluations and a final evaluation of their performance in the scope of 3D printing. As an outcome, a new method for synthesizing functional admixtures for 3D printing applications has been proposed, exhibiting the potential to produce various complex structures.

## 3. Materials and experimental protocol
### 3.1. Materials

To produce 3DPC, a ternary system composed of i) ordinary Portland cement (OPC) CEM I 42.5R (CEMEX, Poland), ii) silica fume (SF) (Mikrosilika Trade Company), and iii) fly ash (FA) obtained from the Dolna Odra power plant (Poland) was used. Natural river sand (SA) obtained from SKSM, Poland, was used and sieved prior to testing to achieve a ≤2 mm nominal size gradation. The helium pycnometer was used to determine the specific gravities of OPC, FA, and SF, which were found to be 3.07 g/cm$^3$, 2.28 g/cm$^3$, and 2.22 g/cm$^3$, respectively. A fixed amount of high-performance powder polycarboxylate superplasticizer Sika ViscoCrete 111 (SP) was used in all mixes.

### 3.2. Development and synthesis of radiation shielding admixtures

This study evaluated three approaches to incorporate radiation shielding admixtures into 3DPC. As pristine materials, commercially available bismuth oxide ($Bi_2O_3$—cat. No. 637017) and gadolinium oxide ($Gd_2O_3$—cat. No. 278513) purchased from Merck, Poland, respectively, were used. A fixed mass ratio of $Bi_2O_3$ to $Gd_2O_3$ of 1:1 was used for comparison. The density of pristine particles determined using a helium pycnometer (Ultrapyc 1200e, Quantachrome Instruments, Anton Paar) was 8.68 g/cm$^3$, 7.28 g/cm$^3$ (Table 1). In the first approach, a bulk mixture of pristine $Bi_2O_3$/$Gd_2O_3$ particles (density of 7.94 g/cm$^3$) was used, and 3DPC specimens were designated as BG. In two other approaches, using a sol-gel method, $Bi_2O_3$/$Gd_2O_3$ particles (with a 1:1 mass ratio) were coated with a silica shell ($SiO_2$). Depending on the synthesis method (sample designation A or B), $Bi_2O_3$-$Gd_2O_3$-$SiO_2$ structures varied with morphology, specific surface area, and density were produced (Table 1). The synthesis methods of structures with comprehensive chemical/physical evaluations were reported in our previous work of Cendrowski et al. [31]. Method A is a typical sol-gel method proposed by Stöber for synthesizing solid silica structures. This procedure is based on the synthesis of silica using an ammonia-ethanol solvent, while in proposed method B, distilled water considerably replaced ethanol in the mix. Synthesis of $SiO_2$ *via* method B enables a reduction of 80% of the ethanol and thus dramatically decreases the final costs of the synthesized material—however, such an approach results in the production of $SiO_2$ with different morphology [25]. The schematic synthesis process of particles is presented in Figure 1. A difference between specimens can be found in their physical and chemical properties.

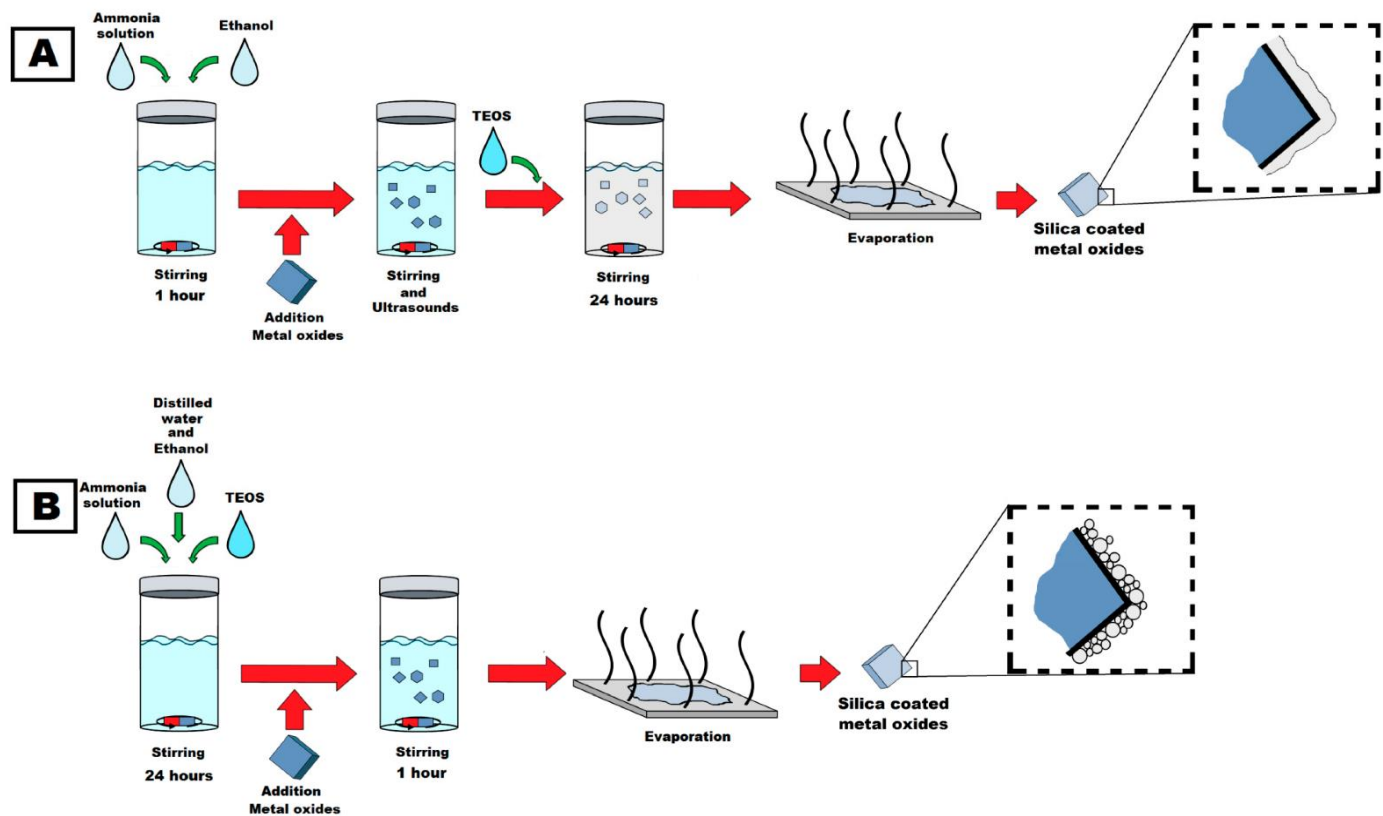

Figure 1. Graphical concept of admixtures production using synthesis methods A and B [31].

The developed particles' specific surface area and total pore volume were determined based on the low temperature (-196 °C) nitrogen adsorption (see Table 1). It is visible that the materials modified using silica possessed a higher specific surface area than the pristine (reference) samples. The specific surface area for $Bi_2O_3/Gd_2O_3/SiO_2$—A material was 8.01 m$^2$/g, whereas for $Bi_2O_3/Gd_2O_3/SiO_2$—B, it was 45.02 m$^2$/g, respectively. The value of the total pore volume increased from 0.013 cm$^3$/g for the $Bi_2O_3/Gd_2O_3$ mixture up to 0.047 cm$^3$/g for $Bi_2O_3/Gd_2O_3/SiO_2$—B material. In contrast, coating using method A did not increase the total pore volume compared to the reference material. The total pore volume of $Bi_2O_3/Gd_2O_3/SiO_2$—A was equal to 0.002 cm$^3$/g, and it was smaller to the value obtained for the reference material $Bi_2O_3/Gd_2O_3$ was equal to 0.013 cm$^3$/g.

Table 1. Density and properties of structures determined from BET. Based on [31]

| Sample | Density [g/cm$^3$] | Specific Surface Area [m$^2$/g] | Total Pore Volume [cm$^3$/g] |
| --- | --- | --- | --- |
| Bi$_2$O$_3$ | 8.68 | 1.10 | <0.001 |
| Gd$_2$O$_3$ | 7.28 | 0.17 | <0.001 |
| Bi$_2$O$_3$/Gd$_2$O$_3$ | 7.94 | 5.20 | 0.013 |
| Bi$_2$O$_3$/Gd$_2$O$_3$/SiO$_2$—A | 5.46 | 8.01 | 0.002 |
| Bi$_2$O$_3$/Gd$_2$O$_3$/SiO$_2$—B | 5.35 | 45.02 | 0.047 |

The specific surface area and the total pore volume results prove that modifying metal oxides with silica was successful. The mixture of pristine Bi$_2$O$_3$/Gd$_2$O$_3$ particles (due to their different sizes) creates a more porous structure than the Bi$_2$O$_3$ and Gd$_2$O$_3$ analyzed separately. The higher pore volume of the metal oxide mixture resulted in the increased specific surface area. In the sample, Bi$_2$O$_3$/Gd$_2$O$_3$/SiO$_2$—A, after silica coating, the pores present in the Bi$_2$O$_3$/Gd$_2$O$_3$ mixture were neutralized/filled by the silica. However, since silica has a significantly lower density than metal oxides, the specific surface area increased further. In the case of Bi$_2$O$_3$/Gd$_2$O$_3$/SiO$_2$—B, silica coating did not only increase specific surface area but also total pore volume. It suggests that formed silica has some porous structure compared to the solid silica synthesized according to the Stöber procedure (method A) [32].

The specific surface area and the total pore volume analyses agree with the TEM and EDS analysis. The TEM images of Bi$_2$O$_3$/Gd$_2$O$_3$/SiO$_2$—A prove that the material comprises a thin silica coating and different-sized metal oxide particles (Figure 2 a-c). These particles can be divided into two groups: (i) particles with elliptical shape, in ranges from 93 nm to 643 nm, with a mean diameter of about 250 nm, (ii) and particles with a more spherical and regular shape, in crystals size range from 101 nm to 248 nm. The particles from groups (i) and (ii) are identified as bismuth oxide and gadolinium oxide, respectively. Detailed analysis of the pristine metal oxide particles was reported previously [31]. The mean thickness of the silica shell was estimated from TEM images to be around 30 nm. High-magnification TEM images of the silica coating prove their solid structure. Previous studies have verified that the silica synthesized according to method A had a solid structure [31]. Contrary to the silica coating synthesized using method A, Bi$_2$O$_3$/Gd$_2$O$_3$/SiO$_2$—B coating shows porous structures with irregular thickness readings from 10 nm to 50 nm. TEM images of the Bi$_2$O$_3$/Gd$_2$O$_3$/SiO$_2$—B are presented in Figure 2 d-f. The differences between silica coatings (structure, size, color) are noticeable in samples lacking metal oxides [25,31].

The process of metal oxide coating with silica did not affect the crystal structure of the Bi$_2$O$_3$ and Gd$_2$O$_3$. All peaks in the samples before and after coating were assigned to the β-Bi$_2$O$_3$ and Gd$_2$O$_3$ phase, according to the PDF 27-0050 and JCPDS card no. 43-1014, respectively. XRD spectra of metals with and without silica coating are presented in Figure 2 g [31]. Previous studies on the separately coated Bi$_2$O$_3$ and Gd$_2$O$_3$ prove that silica coating does not affect metal oxide crystal structure [25].

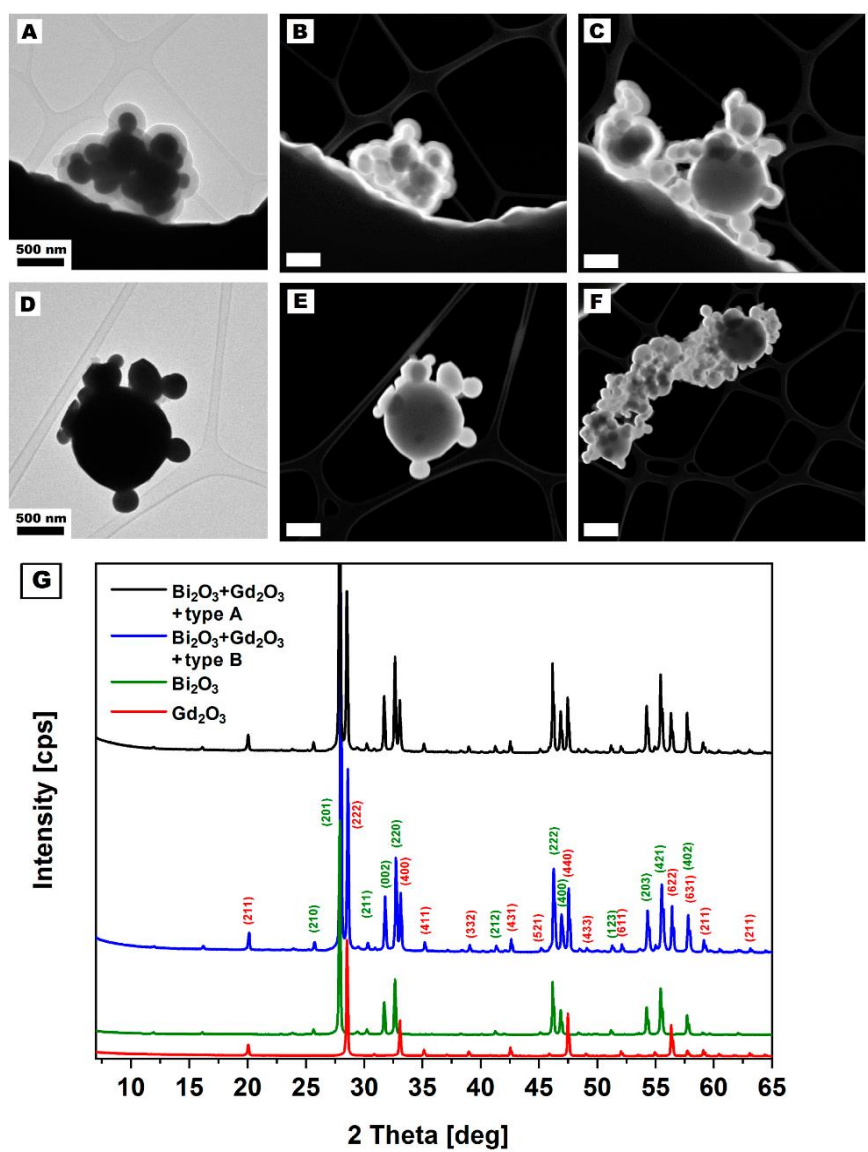

Figure 2. TEM and STEM images of Bi$_2$O$_3$ and Gd$_2$O$_3$ covered with silica synthesized via methods A (A–C) and B (D–F). The white bar on the STEM images corresponds to the length of 500 nm. XRD spectrum

(G) shows signals from the gadolinium and bismuth oxide coated with the silica (*via* different methods) and pristine metal oxides. Reprinted from [31]

### 3.3. Mixture composition and mixing procedure

A total of ten 3DPC mixes were produced (Table 2). The reference (control) mix was designated as C with w/b=0.25. The water content in all mixes remained fixed. In modified mixes, cement has been replaced by volume (vol%) with synthesized structures. Mixes containing pristine nano-sized $Bi_2O_3$ and $Gd_2O_3$ structures were designed as BG, while specimens containing silica-coated structures synthesized using methods A and B were designated A and B, respectively. Digits 125, 250, and 500 describe cement's volumetric replacement with particles BG, A, and B by 1.25, 2.50, and 5.0%.

Table 2. Composition of 3DPC [kg/m$^3$]

| Material | CEM I 42.5R | FA | SF | $Bi_2O_3/Gd_2O_3$ | $Bi_2O_3/Gd_2O_3$-coating A | $Bi_2O_3/Gd_2O_3$-coating B | Sand | Water | SP |
|---|---|---|---|---|---|---|---|---|---|
| Control (C) | 580.0 | 166 | 83 | - | - | - | 1300 | 205 | 2.3 |
| BG125 | 572.8 | 166 | 83 | 18.3 | - | - | 1300 | 205 | 2.3 |
| BG250 | 565.5 | 166 | 83 | 36.7 | - | - | 1300 | 205 | 2.3 |
| BG500 | 551.0 | 166 | 83 | 73.3 | - | - | 1300 | 205 | 2.3 |
| A125 | 572.8 | 166 | 83 | - | 12.6 | - | 1300 | 205 | 2.3 |
| A250 | 565.5 | 166 | 83 | - | 25.2 | - | 1300 | 205 | 2.3 |
| A500 | 551.0 | 166 | 83 | - | 50.4 | - | 1300 | 205 | 2.3 |
| B120 | 572.8 | 166 | 83 | - | - | 12.4 | 1300 | 205 | 2.3 |
| B250 | 565.5 | 166 | 83 | - | - | 24.7 | 1300 | 205 | 2.3 |
| B500 | 551.0 | 166 | 83 | - | - | 49.4 | 1300 | 205 | 2.3 |

In order to produce 3DPC, the following mixing protocol has been applied: (1) slow mixing of dry components for 120 s, (2) addition of water (C sample) or solution with admixtures (ultrasonicated for 10 min using ultrasonic bath), (3) mixing for 540 s. Depending on the use, specimens were either cast or manually transported to an extruder for printing and testing.

### 3.4. Experimental protocol

The experimental protocol has been divided into two main phases and is depicted in Figure 3. In Phase 1, all ten mixtures were pre-screened for their potential applicability in 3D printing applications based on the hydration, rheological, and mechanical studies. In Phase 2 selected mixes were then specifically tested for 3D printing and radiation shielding performance.

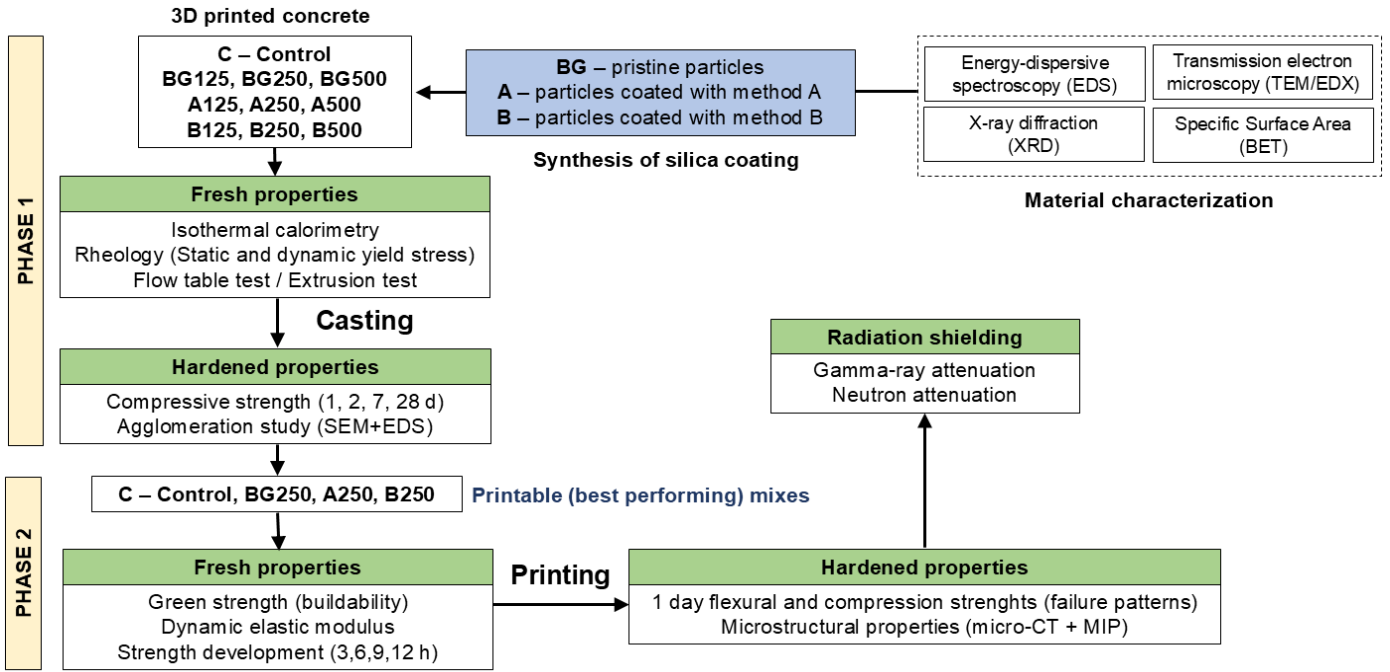

Figure 3. Summary of experimental protocol

## 3.5. Methods

### 3.5.1. Isothermal calorimetry

Isothermal calorimetry was performed on the binder phase (excluding aggregate) of 3DPC using a TAM Air 8-channel calorimeter (TA Instruments, USA) at 20 °C for 168 h. The material was mixed externally using a laboratory mixer, and afterward, 12.96 g of material was placed immediately in a 20 mL PE vial and inserted into the calorimeter.

### 3.5.2. Consistency and extrudability test

In order to evaluate the consistency of mixtures, a flow table test conforming EN 1015-3 was carried out. The measurements were performed at specific intervals (15, 30, 45, and 60 minutes) after introducing water to the dry components ($T_0$). Additionally, specimens C, B250, and B500 were extruded using a 3D printer in various time intervals to verify the flow table observations and confirm that the flow table can be successfully used to pre-screen the potential printability of mixtures.

### 3.5.3. Rheological properties

The rheological measurements were carried out using the MCR 72 (Anton Paar) compact rheometer with vane geometry. A ribbed cup with a volume of 120 cm$^3$ was employed to prevent a wall slip effect. To ascertain the static yield stress ($\tau_s$) across various temporal intervals, the peak value ($\tau_{max}$) derived from the shear stress curve was employed for analysis. The static yield stress was evaluated five minutes after the mixing process. Initially, the mixtures were subjected to a pre-shearing process at a rate of 100 s$^{-1}$ for 60 s, followed by a four-minute resting period. Subsequently, a uniform shear rate of 0.1 s$^{-1}$ was applied for 60 s, with an additional four-minute rest interval. The preceding stage was iteratively replicated at five-minute intervals, extending up to 20 minutes post-mixing.

The evaluation of dynamic yield stress commenced five minutes after the finish of the mixing process. First, the mixture underwent a pre-shearing process for 60 s at a constant shear rate of 100 s$^{-1}$. Subsequently, the shear stress of the mixtures was ascertained over 232 s at 31 designated points distributed logarithmically. The experiment involved a range of shear rates from 100 s$^{-1}$ to 0.1 s$^{-1}$. The Herschel-Bulkley model (HBM) was selected in this study as it demonstrated the highest degree of accuracy in conforming to the dataset, achieving a minimal correlation coefficient ($R^2$) of 0.98542.

The assessment of thixotropy in this study involved calculating the hysteresis loop area. The thixotropy was determined by analyzing the area between the up and down curves of the loop. The thixotropic loop, enclosed by the up and down curves, indicates the deterioration of the paste structures. The Eq. (1) can be utilized to quantify the area of the thixotropic loop.

$$A = \int_{\dot{\gamma}_{min}}^{\dot{\gamma}_{max}} \eta_1 \dot{\gamma}\, d\gamma - \int_{\dot{\gamma}_{min}}^{\dot{\gamma}_{max}} \eta_2 \dot{\gamma}\, d\gamma \qquad \text{Eq. (1)}$$

where: *A* is the area of the thixotropic loop, is the shear rate, and represents the up and down apparent viscosity.

### 3.5.4. *Compressive strength testing*

Initial screening of the compressive strength of mixes, was determined on cast cubic specimens (40 × 40 × 40 mm) using electromechanical universal tester UNIFRAME 250 (CONTROLS S.p.A., Liscate, Italy) with a loading rate of 0.6 MPa/s. The measurement was performed after 1 d, 2 d, 7 d, and 28 d of curing. Five specimens for each composition were tested, and the mean value was taken as a representative. In the Phase 2 of the experimental protocol, selected mixes (i.e., C, BG250, A250, B250) were evaluated additionally in the scope of early compressive strength build-up. Compressive strength was tested after 3, 6, 9, and 12 h on three cubic specimens (40 × 40 × 40 mm) using a ToniTechnik hydraulic press machine (at 10/25/50/100 N/s, for each consecutive period of testing).

### 3.5.5. Green strength (buildability test)

The uniaxial unconfined compressive test (UUCT) was performed on cylindrical specimens with D=60 mm and H=120 mm dimensions. The specimen preparation procedures were the same as those for printed specimens, described in detail in point 3.5.7. Subsequently, the mixing process lasted for about 9 min. After this duration, the samples were molded into plastic cylindrical forms. Prior to commencing the test, the specimens were de-molded. The tests were carried out after 30 and 60 min after water contact. The maximum stresses ("green" strength) and the green Young Modulus were evaluated using the procedure presented in [33].

### 3.5.6. Dynamic elastic modulus

The development of dynamic elastic modulus in the first 24 h of hydration was determined using a Vikasonic (Schleibinger) ultrasonic device. Fresh 3DPC was poured into a dedicated Vicat ring, and measurement was performed by placing the specimen between two ultrasonic transducers supplied with 1 s or less pulses per second at 54 kHz.

### 3.5.7. Printed specimens preparation

The specimens were printed using a 3 degrees of freedom (3DoF) Cartesian robot (gantry). The printing process was performed using a special printing head/extrusion system with an auger and a 40 × 8 mm nozzle at a 45° angle. A detailed description of the mentioned head is presented in another work [33]. The materials were delivered directly to the head. The interior auger controlled the mix's flow and ensured the printed material's appropriate quality. An example of the printing process is presented in Figure 4a. This study used a printing rate of 4500 mm/min.

The mixing process consisted of the following steps: 1) the binder was mixed with water and superplasticizer in a two-stage regime: i) low speed (54 rpm) for 1 minute, and ii) high speed (108 rpm) for 2 minutes. The aggregate was added to the mix, and another two-stage mixing was performed: i) low speed (54 rpm) for 2 minutes and ii) high speed (108 rpm) for 4 minutes.

During the printing process, the following types of specimens were made:

1) Linear 5-layered (height of 8 mm and a width of 40 mm of each layer) specimens for 1-day flexural and compression strength tests. Before testing, the specimens were cut into 160 mm long pieces (Figure 4 b). These specimens were used to determine the failure pattern during the flexural strength test and to evaluate compressive strength.
2) Specimens for evaluating interlayer adhesion. These specimens consisted of 4 layers and were evaluated using a splitting test.
3) Small specimens for microstructural properties. Cut from linear specimens (described in the point above) into 20 mm long elements.

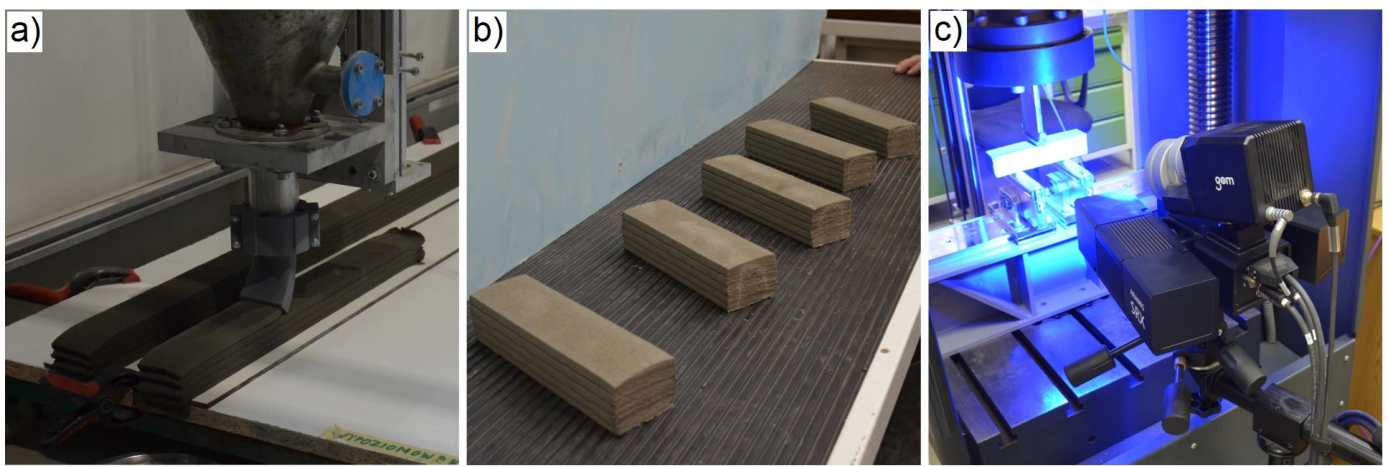

Figure 4. Example of printing process (a), printed specimens (b), and specimens during flexural strength test with DIC system (c)

### 3.5.8. Failure pattern evaluation during mechanical tests

The flexural and compressive strength of the printed specimens were evaluated at 1 day. The test has two main aims: 1) to compare findings with casted specimens and 2) to assess the impact of mix composition on the failure pattern of multilayered specimens during the flexural test. The compressive strength test was performed on printed specimens using the procedure described in point 3.5.4. The flexural strength test was performed according to the EN 1015-11 on 40 mm × 40 mm × 160 mm cast specimens and five layered printed specimens with the exact dimensions using a Zwick/Roell Z600 electromechanical press. The tests were performed in a direction perpendicular to the printing layers to evaluate the failure pattern of 3D printed samples due to a possible reduction of interlayer adhesion. The digital image correlation method (DIC) ARAMIS SRX (GOM) was used to analyze the failure pattern. The MV300 measurement space was used in this study. It consists of two 4K cameras set at an angle of 25° to each other on a 180 mm long measuring beam. A particular black-and-white pattern was prepared on the side surfaces of the samples, creating a surface component - an element enabling the display of a map of displacements or deformations for a selected surface of the element. The example of specimens during the test is presented in Figure 4c. Analysis was conducted using GOM Correlate software (ARAMIS system) which allowed to analyse the failure pattern of specimens, including the force–deformation relationship.

### 3.5.9. Interlayer adhesion evaluation

The splitting test method evaluated the interlayer adhesion according to EN 12390-6 and similar to the other research focused on 3D concrete printing [34,35]. The tests were performed on specimen b × h × L = 40 ± 5 mm × 40 ± 5 mm × 40 mm. The final results were evaluated as splitting tensile force ($f_{cl}$ [MPa]) according to the EN 12390-6 (Eq. 2):

$$f_{cl} = \frac{2F}{\pi \cdot l \cdot d} \quad \text{Eq. (2)}$$

where $F$ is maximum force recorded by an electromechanical press [N], $l$ is the length of the line contact of the specimen [mm], and $d$ is the designated cross-sectional dimensions [mm].

*3.5.10. Microstructural investigations (SEM+EDS, MIP, micro-CT)*

A Hitachi TM3000 SEM instrument with Bruker Quantax EDS was used to assess the uniformity of the distribution and the possible occurrence of coagulation of silica-coated bismuth and gadolinium oxides. Specimens with an area of approx. 2 cm$^2$ and a thickness of up to 3 mm were prepared for the tests from the middle sections of cubic specimens. The samples were sputtered with gold-palladium alloy in a vacuum chamber to increase the electrical conductivity of the tested surface. Photos were taken in a magnification range from 50× to 1000×. Energy-dispersive X-ray spectroscopy was used to assess the elemental composition and the distribution of the bismuth element in the specimen. Before the EDS tests, the device was calibrated on a reference copper specimen.

The mercury intrusion porosimetry tests were performed using Quantachrome Poremaster 60. The tests were performed after 28 days of curing. The specimens in question (7 mm × 0.7 mm × 20 mm) were cut from the middle section of cubic samples 40×40×40 mm. For each composite, two tests were conducted to verify the repeatability of the results. Before the tests, the specimens were dried at 70 °C for 48 hours. The mercury surface tension was set at 0.48 N/m, and the contact angle was set to 140° upon intrusion. The specimens were placed into measurement cells and filled with mercury in a low-pressure chamber (to 0.34 MPa). Furthermore, the cells were inserted into a pressure chamber and subjected to high pressures (up to 413 MPa). Based on the tests, cumulative and log differential graphs of porosity distribution were made, and the basic properties of composites were given, such as total porosity, specific surface area, and volume density.

X-ray micro-computed tomography (micro-CT) was utilized for selected specimens: C (control), BG250, A250, and B250. Each specimen was cut into 2.0 cm cubes for an optimum image resolution in the micro-CT investigation. Initially, scanned images were reconstructed and processed to segment the pore and solid regions for a more effective investigation. Then, the binarized images can be obtained, and a 3D data structure can be generated by subsequent stacking of the segmented images. The used image resolution was 800 × 600 × 500 voxels for *xx yxz* directions, where the *z*-direction is of specimen height. Each voxel has a size of 19.97 μm, which is enough to describe voids that can affect the mechanical and thermal properties of the materials. The pore characteristics, such as porosity, pore size distribution, and sphericity, were evaluated using the binarized 3D volume of each specimen. The homogeneity of the solid structures was also confirmed using the micro-CT data, and they were correlated with the material properties.

### 3.5.11. Radiation shielding performance

The radiation shielding effectiveness of the synthesized samples was evaluated using narrow-beam transmission geometry. A detailed description of the experimental setup has been reported in the authors' previous publication [36]. The gamma-ray attenuation experiments utilized collimated sources of gamma rays emitting various energies. Specifically, radioactive point sources of $^{133}$Ba, $^{137}$Cs, $^{60}$Co, and $^{233}$Th were employed, covering an energy range of 81 keV to 2614 keV. These sources served a dual purpose: irradiating the samples and calibrating the detector. The transmitted radiation was recorded and analyzed using a 3″ × 3″ NaI(Tl) scintillator detector coupled with a multichannel analyzer and Genie 2000 software (Canberra). An investigation of the slow neutron shielding properties was also conducted. This experiment utilized an Am-Be neutron source with an activity of 3.7 GBq.

Various key parameters that measure the material's ability to attenuate radiation were determined based on fine beam geometric configuration principles and the Lambert-Beer Law. These parameters include the linear attenuation coefficient (LAC, μ in cm$^{-1}$), total macroscopic cross-section for slow neutron attenuation (Σs in cm$^{-1}$), half-value layer (HVL in cm), and mean free path (MFP in cm). The equations for these calculations are detailed in Table 3.

Table 3. Equations used to determine gamma ray and neutron shielding parameters

| Parameter | Symbol | Unit | Equation | Explanation |
| --- | --- | --- | --- | --- |
| Linear attenuation coefficient (LAC) | $\mu$ | cm$^{-1}$ | $\mu = -\frac{1}{x}\ln(I/I_0)$ | $x$: sample thickness. $I_0$(incident) and I(transmitted) photon intensities. |
| Macroscopic slow neutron cross-section | $\Sigma_s$ | cm$^2$/g | $\Sigma_s = -\frac{1}{x}\ln(\phi/\phi_0)$ | $\phi_0$(incident) and $\phi$ (transmitted) neutron fluxes |
| Half-value layer | HVL | cm | $HVL = \ln2/\Sigma_s;\ \ln2/\mu$ | |
| Mean free path | MFP | cm | $MFP = 1/\Sigma_s;\ 1/\mu$ | |

# 4. Results and discussion

## 4.1. Isothermal calorimetry

Figure 5 presents the results of the heat flow of the binder phase (no aggregate) of 3DPC with varied particle addition. Apparent differences in the heat flow can be distinguished depending on the type of structures, and their content can be observed. As expected, the C specimen exhibited the fastest hydration with the highest power of exothermic peak, which can be distinguished. Introducing BG particles dramatically decreased the peak magnitude and occurrence (Table 4). BG125, BG250, and BG500 exhibited 26, 28, and 38 % decrement of power of exothermic peaks when compared to C, along with noticeable retardation of the peak occurrence reaching up to 48% (BG500). This effect is attributed to retarding effect of $Bi_2O_3$ in the cementitious systems. As reported by previous research, $Bi_2O_3$ dramatically affects the hydration process, and this effect seems to increase along with the decrease in particle size [24,25,36]. On the contrary, only very few authors discussed the effects of $Gd_2O_3$ on the hydration process; however, it was reported that this material has a relatively negligible or could even lead to a slight acceleration of the hydration process [25,37,38]. The introduction of silica-coated structures resulted in an improvement in the performance of structures in the binder system and enabled to reduce of the negative effect of $Bi_2O_3$ presence. First, specimens containing structures coated in A-type silica showed the highest reactivity among other mixes. This is attributed to the high amorphicity of the silica coating synthesized in purely ethanol solvent. In this case, in both A125 and A250, the power of exothermic peaks has been decreased only by 8% when compared to C. Slightly lower power of exothermic peaks was found in B-type coated specimens where silica coating was found to be less reactive because of other morphological features of the material as a result of introduction of distilled water to solvent. This phenomenon was comprehensively described in our previous work [25]. In this case, B125 and B250 exhibited 12% and 20% lower power of exothermic peaks when compared to C. What is worth noting is that regardless of coating type, the occurrence of exothermic peaks was comparable regardless of synthesis method and particle content. Therefore, introducing silica-coated structures reduced the delay of exothermic peak occurrence below 10% in all specimens (A and B) compared to the Control sample. Moreover, in almost all cases, regardless of content (excluding B500), all specimens exhibited higher power of exothermic peak than BG125 (for B500 values were comparable). At the same time, introducing pristine BG particles resulted in a noticeable decrement in the cumulative heat value after seven days, reaching up to 16% when compared to the C. On the contrary, specimens of type A and B exhibited only minimally lower cumulative heat values when compared to C. Therefore, the introduction of A or B-type silica-coating enabled to overcome the retarding effect of $Bi_2O_3$, and thus the occurrence of peaks was only slightly delayed when compared to Control (C) specimen. At the same time, cumulative heat values after seven days also remained similar to those of the C specimen.

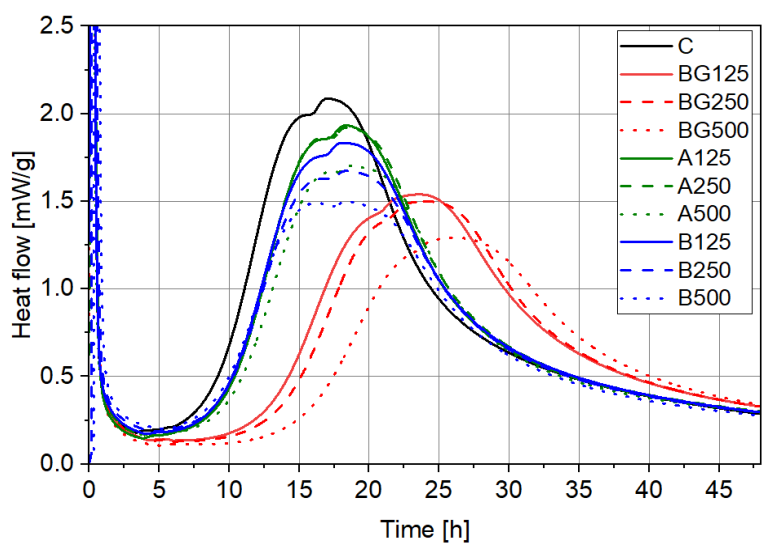

Figure 5. Heat flow in the first 48 hours of hydration

Table 4. Comparison of maximum heat and cumulative heat of 3DPC binder phase

| Material | Maximum heat [mW/g] | Loss in comparison to C [%] | Peak occurrence [h] | Loss in comparison to C [%] | Cumulative heat after 168 h [J/g] | Loss in comparison to C [%] |
|---|---|---|---|---|---|---|
| Control (C) | 2.09 | - | 16 h 53 min | - | 210 | - |
| BG125 | 1.54 | -26 | 22 h 52 min | -36 | 195 | -7 |
| BG250 | 1.50 | -28 | 23 h 13 min | -38 | 191 | -9 |
| BG500 | 1.29 | -38 | 24 h 58 min | -48 | 176 | -16 |
| A125 | 1.93 | -8 | 18 h 01 min | -7 | 202 | -4 |
| A250 | 1.92 | -8 | 18 h 06 min | -7 | 208 | -1 |
| A500 | 1.70 | -19 | 18 h 23 min | -9 | 202 | -4 |
| B125 | 1.83 | -12 | 17 h 44 min | -5 | 204 | -3 |
| B250 | 1.67 | -20 | 17 h 58 min | -6 | 202 | -4 |
| B500 | 1.50 | -28 | 18 h 07 min | -7 | 196 | -7 |

**4.2. Rheological properties**

*4.2.1. Consistency and extrudability*

Table 5 presents the changes in flowability of all the developed mixtures at 15min intervals up to 60 min post-mixing. The average spread diameter of the fresh mixtures in each time interval was measured to examine the flow development of the mixtures over time. The measurements allowed us to investigate the impact of the BG particle incorporation and silica coating methods on the mixtures' flowability. Based on available state-of-art reports [39,40], it has been determined that the mixtures with an average spread diameter falling within the ranges of 185-145 mm, 140-135 mm, and < 135mm can be classified as "printable", "potential threats with extrusion", and "not printable", respectively.

The spread diameter on the flow table test is significantly affected by the incorporation of BG particles regardless of whether these particles are silica-coated or not. Including plain BG particles with and without silica coatings reduces the 3D printable window. The silica-coated BG particles in the mixture absorbed more water due to their high volume of particles and increased specific surface area. As a result, the effective water content in the mixture decreased, reducing the spread diameter. Moreover, a high concentration of particles caused particle aggregation, resulting in a significant decrease in the spread diameter of mixtures containing BG, A, or B particles. These were confirmed by scanning electron microscope analysis in Section 4.4.

The results indicated that the mixtures containing 5 vol% of non-coated and coated BG particles (i.e., BG500, A500, and B500) exhibited the lowest speed diameter (i.e., approximately < 140 mm), suggesting a printing window of less than 15 min, which is not applicable to use as a feedstock in additive manufacturing technology. Confirmation of the results obtained from the flow table test was achieved by implementing the open extrusion window test (Figure 6). It confirmed that the most effective dosage of incorporating silica-coated BG particles is 2.5 vol%. For example, based on the capacity and power of the 3D printing setup utilized in this study (as illustrated in Figure 6), the B250 blend exhibited an ideal open extrusion window of 40 min. However, 5 vol% BG particles coated via method B (i.e., B500) reduced the open extrusion window to less than 15 min, a data line reported during flow table test measurement.

Table 5. Flow table test results with extrudability color indication

| Sample | Spread flow [mm]* | | | |
|---|---|---|---|---|
| | 15 min | 30 min | 45 min | 60 min |
| Control (C) | 163 | 154 | 147 | 145 |
| BG125 | 168 | 152 | 146 | 145 |
| BG250 | 164 | 149 | 138 | 132 |
| BG500 | 142 | 137 | 132 | 128 |
| A125 | 154 | 149 | 146 | 142 |
| A250 | 154 | 146 | 141 | 135 |
| A500 | 135 | 130 | 124 | 120 |
| B125 | 163 | 154 | 144 | 140 |
| B250 | 153 | 147 | 135 | 131 |
| B500 | 137 | 128 | 120 | 115 |

*Green - 185-145 mm (printable), orange – 135-140 mm (potential threats with extrusion), red - <135 mm (not printable).

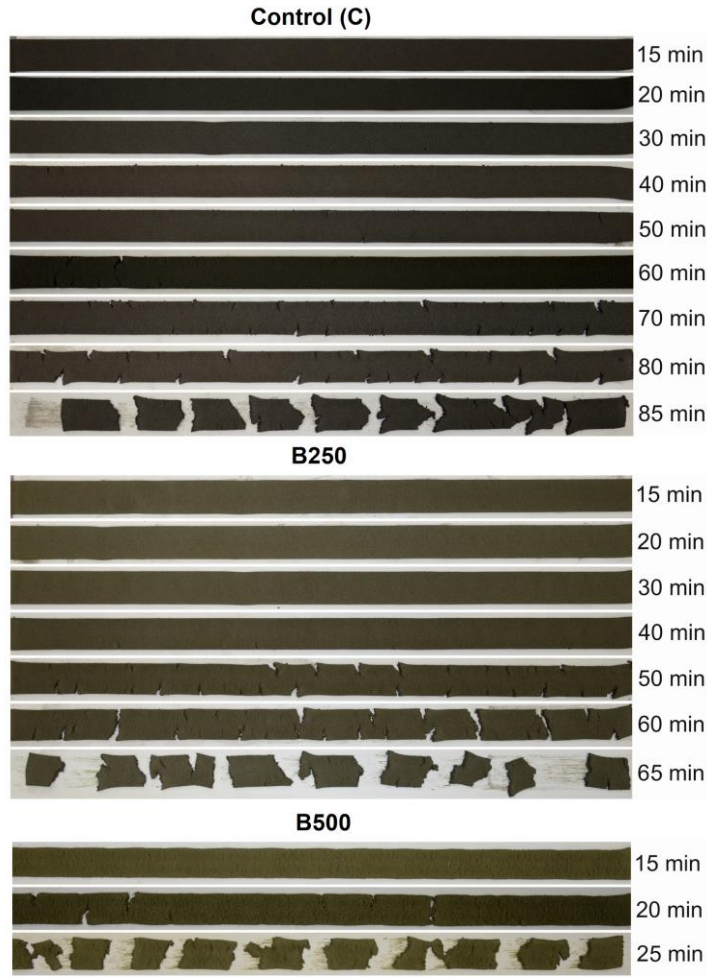

Figure 6. Verification of flow table results through extrusion test (open time) of C, B250, and B500 mixes

### 4.2.2. Static yield stress development

Table 6 display the $\tau_s$ of mixtures with different fractions of BG aggregates, both with and without coatings. The results demonstrate a gradual increase in $\tau_s$ for all mix categories as testing time increases. Notably, incorporating BG particles led to a significant enhancement in $\tau_s$. For instance, the initial $\tau_s$ (i.e., testing after 5 min) increased from about 245 Pa for M0 to 346 Pa, 360 Pa, and 314 Pa for BG125, BG250, and BG500, respectively. This trend persisted for the $\tau_s$ of the mixtures mentioned above after 10 min, 15 min, and 20 min intervals.

The implementation of coating method A has altered the static yield stress pattern. As shown in Table 6, when utilizing particles coated with method A, a significant increase in the $\tau_s$ was observed for the A125 mixture compared to that of the control sample (i.e., M0). Subsequently, the $\tau_s$ gradually decreased for both A250 and A500, respectively, at each time interval. It should be mentioned that the $\tau_s$ value of the A250 and A500 samples at each time interval is lower than that of their corresponding mixtures containing plain BG particles (i.e., BG250 and BG500). It shows that for these mixtures, the silica coating exhibits a lubrication effect, facilitating the initial flow of the material. The effect observed in the A125 mixture does not conform to the same trend as the other mixtures coated with Method A, which has to be associated with the lower dosage of silica-coated BG particles in the mixtures.

During each time interval, coating method B decreased $\tau_s$ values of the B125 and B250 compared to their uncoated counterparts (i.e., BG125 and BG250). It could be attributed to the lubrication effect of the silica-coated particles in lower dosages under static shear rates. However, the lubrication effect of Method B is less pronounced than Method A. As hypothesized, the lubrication effect may be more dominant under a constant shear rate (static conditions), particularly at lower dosages, leading to lower shear stress as the steady state allows smoother particle movement with less interlocking. For B500, on the other hand, the $\tau_s$ values increased when compared to BG500 at each time interval.

Table 6. Static yield stress values obtained at various time intervals

|       | Static stress (Pa) | | | |
|-------|--------|--------|--------|--------|
|       | 5 min  | 10 min | 15 min | 20 min |
| M0    | 245.13 | 354.17 | 409.55 | 446.46 |
| BG125 | 345.99 | 551.37 | 603.44 | 616.21 |
| BG250 | 359.69 | 551.04 | 615.73 | 660.12 |
| BG500 | 314.39 | 499.7  | 576.29 | 630.6  |
| A125  | 332.04 | 562.66 | 601.92 | 642.00 |
| A250  | 248.93 | 448.38 | 525.55 | 532.29 |
| A500 | 114.61 | 319.52 | 405.72 | 453.43 |
| B125 | 306.05 | 438.34 | 493.85 | 547.96 |
| B250 | 349.01 | 525.12 | 589.01 | 639.28 |
| B500 | 352.94 | 538.8 | 607.49 | 670.95 |

### *4.2.3. Dynamic yield stress and thixotropic area*

Based on the data in Table 7 and Figure 7, the dynamic rheology parameters, i.e., yield shear stress, consistency coefficient, and rheological index, were determined for all the combinations employing the HBM fitting model. The results indicated that the incorporation of plain BG additives leads to a substantial increase in the dynamic yield shear stress of the mixture. As seen in Figure 7, the yield shear stress of the composites progressively increased from 100.7 Pa for the C (control mix) to 214.9 Pa and 272.1 Pa for the BG125 and BG250, respectively. However, incorporating more BG particles (i.e., BG500) reduces dynamic yield stress compared to BG125 and BG250. Nonetheless, the value is still higher than that of C. The observed increase in dynamic yield stress of BG125 and BG250 is due to improved particle packing and interlocking within the matrix, resulting in higher resistance to deformation and flow. Incorporating BG particles significantly amplifies interparticle forces, resulting in higher interactions among the particles within the mixture and a more compact microstructure [31]. On the other hand, the decrease in dynamic yield stress in the BG500 mixture is caused by BG particle agglomeration.

Implementing the silica coatings brings about a discernible change in the dynamic rheological characteristics of the composites. The results suggest that the composite with BG particles coated through method A demonstrated a substantial decline in their dynamic yield shear stress and consistency coefficient compared to their plain BG counterparts. The results indicate that the dynamic yield stress of the A125, A250, and A500 decreased by 5%, 67%, and 85%, respectively, compared to their corresponding uncoated mixtures. The consistency coefficient of the A125 and A500 were also reduced by 28% and 48%; however, a slight increase of about 2% in the A250 consistency coefficient was registered compared to that of BG250. As demonstrated in the authors' previous research [31], the silica coating of BG particles improves their dispersion and prevents aggregation. By implementing method A, the dispersion of BG particles has significantly enhanced, reducing particle aggregation and contributing to a more uniformly distributed paste. According to the TEM results outlined in Section 3.2, BG particles coated using method A displayed smooth surfaces devoid of impurities. It enhanced dispersion and smooth-surfaced particles minimize internal friction and facilitate smoother material flow when subjected to shear force, reducing yield shear stress. The application of silica coating possesses the additional benefit of serving as a lubricant or flow-enhancing additive, attributable to its smooth surface and low friction properties. It may potentially aid in reducing friction and increasing resistance to the flow of the mixtures, thereby leading to a decline in yield shear stress.

Table 7. Summary of yield shear stress ($\tau_0$), flow rheology index (n), consistency (K) coefficient, and thixotropic area determined from rheological measurements

| Mix | $T_0$ | n | k | $R^2$ | Thixotropy |
| --- | --- | --- | --- | --- | --- |
| C | 100.7 | 0.35 | 175.09 | 0.9933 | 5901 |
| BG125 | 214.89 | 0.33 | 216.29 | 0.99263 | 10980 |
| BG250 | 272.14 | 0.40 | 197.133 | 0.99471 | 15661 |
| BG500 | 164.411 | 0.34 | 194.889 | 0.99604 | 11428 |
| A125 | 204.222 | 0.428 | 154.550 | 0.99409 | 11379 |
| A250 | 89.57 | 0.455 | 202.905 | 0.99596 | 17476 |
| A500 | 24.138 | 0.519 | 101.776 | 0.99944 | 7760 |
| B125 | 72.110 | 0.291 | 292.835 | 0.99652 | 8950 |
| B250 | 564.466 | 0.473 | 167.939 | 0.98542 | 18792 |
| B500 | 159.648 | 0.365 | 212.191 | 0.99629 | 15716 |

As noted earlier, silica coating typically provides lubrication in mixtures. However, when BG particles are silica-coated *via* method B, their distinct surface properties (see Figure 2) can foster more significant interparticle interaction. This method may introduce additional forces, such as Van der Waals or electrostatic forces, which enhance particle interlocking and result in higher yield shear stress. As a result, the rheological behavior of mixtures containing these particles can vary depending on the specific case, as these two effects can sometimes contradict each other. The results (Table 7) demonstrated that silica coating via method B results in a remarkable increase in the dynamic yield shear stress of the B250 by 107%, compared to the values registered for the BG250 (i.e., 272.1 Pa). However, the B125 and B500 mixtures exhibited 66% and 9% reductions in their dynamic yield shear stress compared to their plain BG counterparts.

The hysteresis loop is created due to the difference in shear stress between the ascending and descending stages at the same shear rate. To comprehend thixotropy, it is crucial to understand that during the ascending stage, a substantial amount of shear stress is necessary to break down the flocculation structure that originates from cement hydration and the strong intermolecular bonding among the primary components of the mixtures. On the other hand, the descending stage experiences a lower degree of shear stress since the flocculation structure and intermolecular bonding have already been disrupted. These phenomena are fundamental to grasp the concept of thixotropy. According to Chen et al. [41], the thixotropic loop may indicate structural failure, but its recovery degree is limited at high shear rates due to various factors. Therefore, using it only for comparing different samples under identical conditions is advisable rather than as a reliable measure of actual thixotropy.

The values of the thixotropic loop area of mixtures are demonstrated in Figure 7. The results indicated that adding BG particles, both with and without coating, can substantially improve the thixotropy of composites compared to that of C. This is attributed to the flocculation induced by the BG particles due to their nano-sized nature, resulting in a higher thixotropic loop area. As per the results, it can be inferred that the mixtures containing 2.5% BG particles (i.e., BG250, A250, and B250) exhibited the highest levels of thixotropy, regardless of the coating type. Notably, the maximum thixotropy area of approximately 18792.8 was registered for the B250, which is 218% higher than that of C, suggesting that coating method B can favorably affect the mixtures' printing performance.

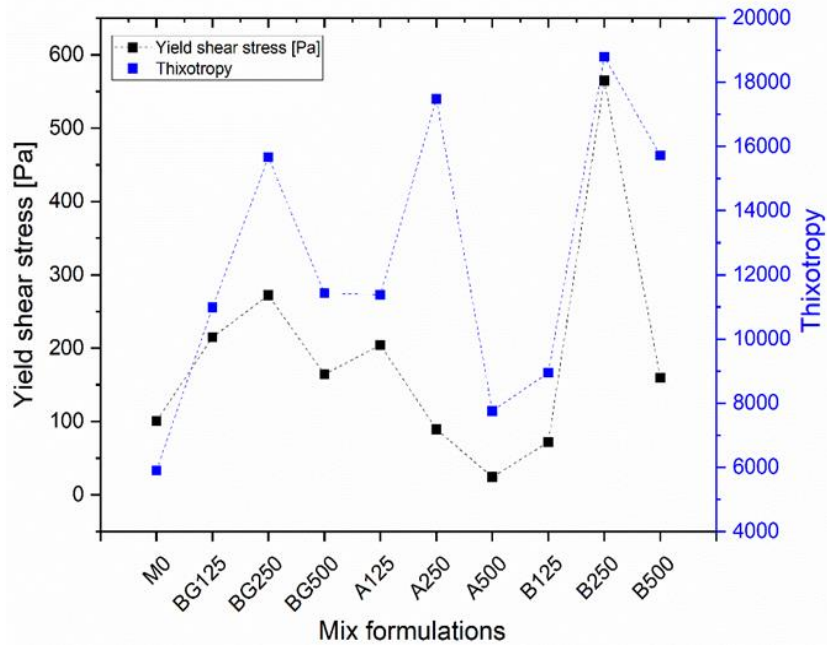

Figure 7. Summary of yield shear stress and thixotropy area calculated for all evaluated mixes

## 4.3. Compressive strength

The results of the compressive strength development are depicted in Figure 8. A discernible impact of the delayed hydration process in the presence of admixtures was observed in the compressive strength of the specimens after one day of curing. The specimens containing pristine particles (BG mixes) showed the lowest compressive strength, while the best-performing mixture was identified as A250. However, the compressive strength of A250 was still lower than that of the C. After 24 hours, the best-performing mixes from each group (BG250, A250, and B250) exhibited compressive strengths 35%, 14%, and 26% lower, respectively, compared to the C. Results are in line with the calorimetric study (Table 4) confirming that particles coated with A-type silica coating exhibit the highest reactivity, which resembles the mechanical performance enhancement. A similar trend can be observed after two days. Previous studies of the authors confirmed that structures coated with silica type A exhibit higher ability in consumption of portlandite (CH) when compared to structures type B at early ages (2 and 7 days); therefore, higher mechanical performance could be expected [31]. After 28 days of curing, the discrepancy between specimens decreased, and specimens exhibited comparable values to those of control specimens. However, the best-performing mixes in all cases contain 2.5 vol% cement replacement with structures. The highest mechanical performance was reported for the B250 (98.7 MPa), followed by Control (95.1 MPa). It can be attributed to optimal packing and homogenous dispersion of the particles within the cementitious matrix. It is well-known in state-of-art that there is an optimal amount of nanomaterials (usually ranging between 1-3%) that can effectively participate in the hydration process through (depending on the type of structure): (1) seeding and (2) pozzolanic reactions along with (3) physical "filling" effect of cement matrix [42]. Excessive amounts of fine admixtures are usually associated with challenges in properly dispersing particles, which is especially pronounced in low w/b systems such as 3DPC or ultra-high performance concretes. Hence, this leads to agglomeration and so-called "weak zones" in the cementitious matrix. Therefore, the inclusion of 5 vol% of particles resulted in difficulties in proper material dispersion within the cementitious matrix, confirmed by scanning electron microscope (SEM) analysis presented in the next section.

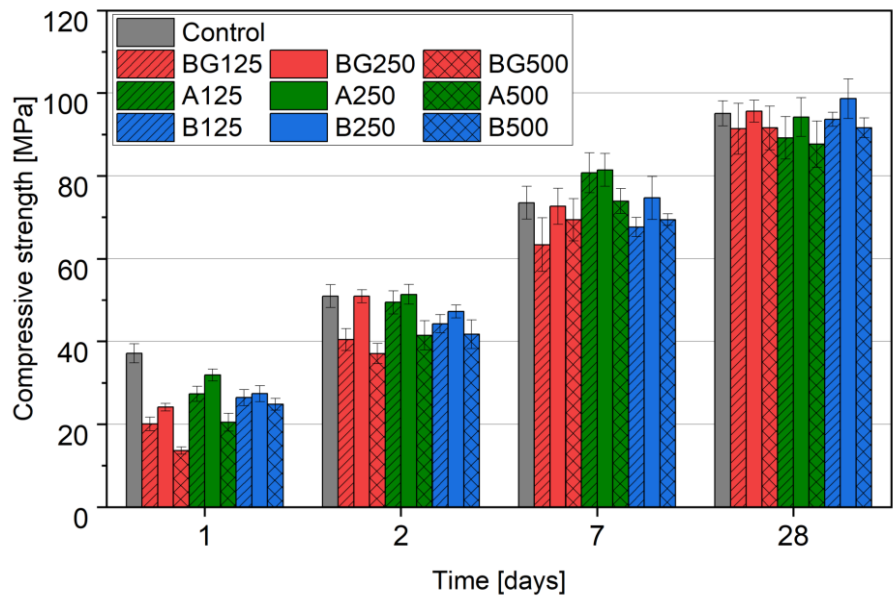

Figure 8. Development of compressive strength of cast 3DPC

Therefore, from the mechanical perspective, it can be concluded that BG specimens exhibited the lowest early mechanical performance followed by B and A structures which are related to the retarding effect of $Bi_2O_3$ in BG specimens as well as particular reactivity of silica shells in A and B mixes. However, after 28 days, no substantial differences between modified and control mixes were found, with the highest performance being 2.5 vol% replacement level in all types of structures.

**4.4. SEM/EDS analysis**

To evaluate the potential agglomeration of the structures within the matrix, an SEM/EDS evaluation was performed to map the Bi signal on the surfaces of selected specimens (Figure 9). For this purpose, specimens containing 2.5 vol% and 5 vol% of structures were selected for evaluation. No signs of large agglomerated materials were observed in specimens containing 2.5 vol% regardless of the type of structures; however, in some areas, a slightly higher signal of Bi was found, confirming the presence of the material. On the contrary, larger agglomerated areas were found in specimens containing 5 vol% of particles, and their intensity was higher in the case of silica-coated structures. As discussed in previous sections, this is attributed to a large volume of particles causing difficulties with ultrasonication due to the limited amount of water available for this process. At the same time, high SSA of silica-coated structures decreases the amount of available water for the cementitious system, which results in difficulties with the dispersion of the material, thus decreasing its consistency (Table 5). Therefore, BG500 exhibited fewer agglomerated areas than A500 and B500.

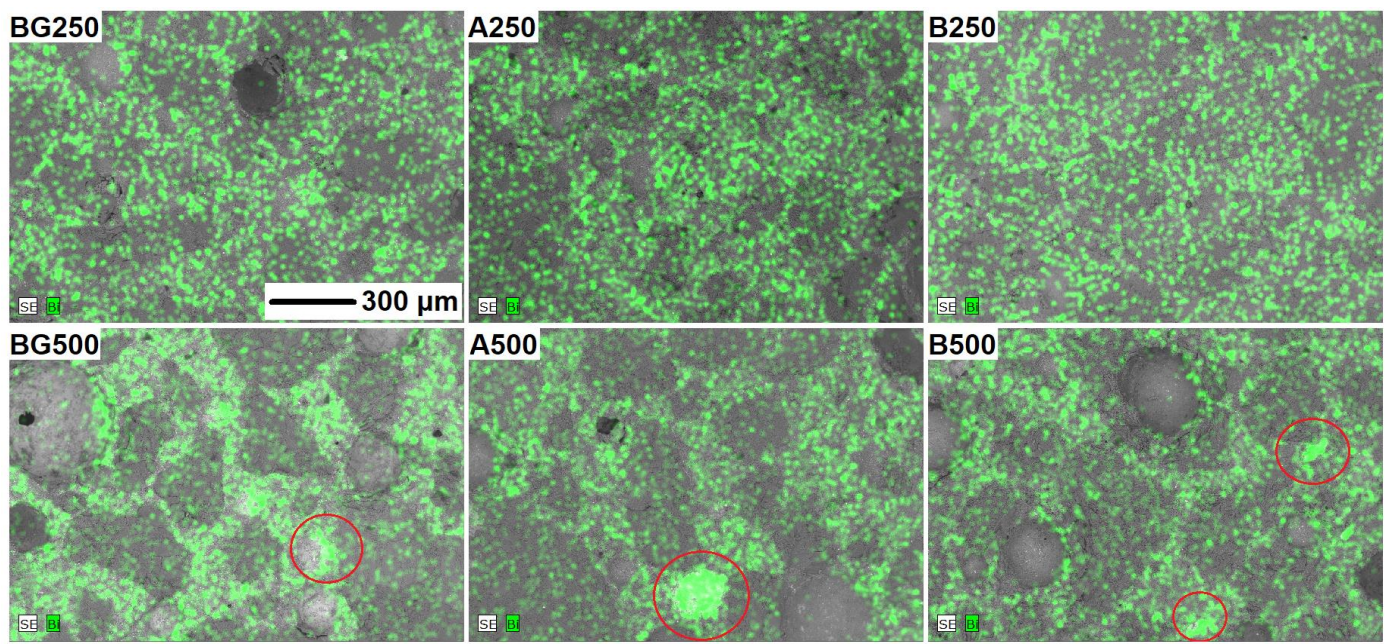

Figure 9. Bi elemental distribution in 3DPC mixes containing 2.5 vol% and 5 vol% of admixtures (scale is unified for all micrographs)

## 4.5. Buildability (green strength) evaluation

Based on the data obtained in previous sections for Phase 2 (Figure 3), it was found that the most suitable mixtures for 3D printing applications contained 2.5 vol% of structures. As a result, mixes BG250, A250, and B250 were selected for further evaluation, along with the Control mix. The comparison of green strength and "green" young Modulus with CoV values is depicted in Figure 10. Adding pristine nano-sized $Bi_2O_3/Gd_2O_3$ particles regardless of coating method type (i.e., A250, B250) benefited green strength and "green" Young's Modulus. The results indicate an increase in green strength values ranging from 19.6% to 62.4% compared to the control sample. Regarding Young's Modulus, the modifications led to an increase ranging from 11.7% to 57.7% compared to the control sample. Young's modulus and green strength were the highest values observed for $Bi_2O_3/Gd_2O_3$ particles coated with method B (B250). Compared to sample C, the green strength of B250 was 62.4% higher after 30 minutes and 46.7% higher after 60 minutes, while Young's modulus was 57.7% higher after 30 minutes and 45.2% higher after 60 minutes. As for mixes BG250 and A250, the increase in green strength ranged between 17.6% and 32.4% compared to the control sample. In addition, modifying pristine nano-sized $Bi_2O_3$ and $Gd_2O_3$ led to an increase in the green strength up to 19.63 kPa after 30 min. A similar relationship exists for Young's Modulus. The presented results show that using pristine nano-sized $Bi_2O_3$ and $Gd_2O_3$ (especially mix B250) has remarkable effects on the buildability of 3D printed mixes, which was later reflected in the pore structure of 3D printed composite (Section 4.7).

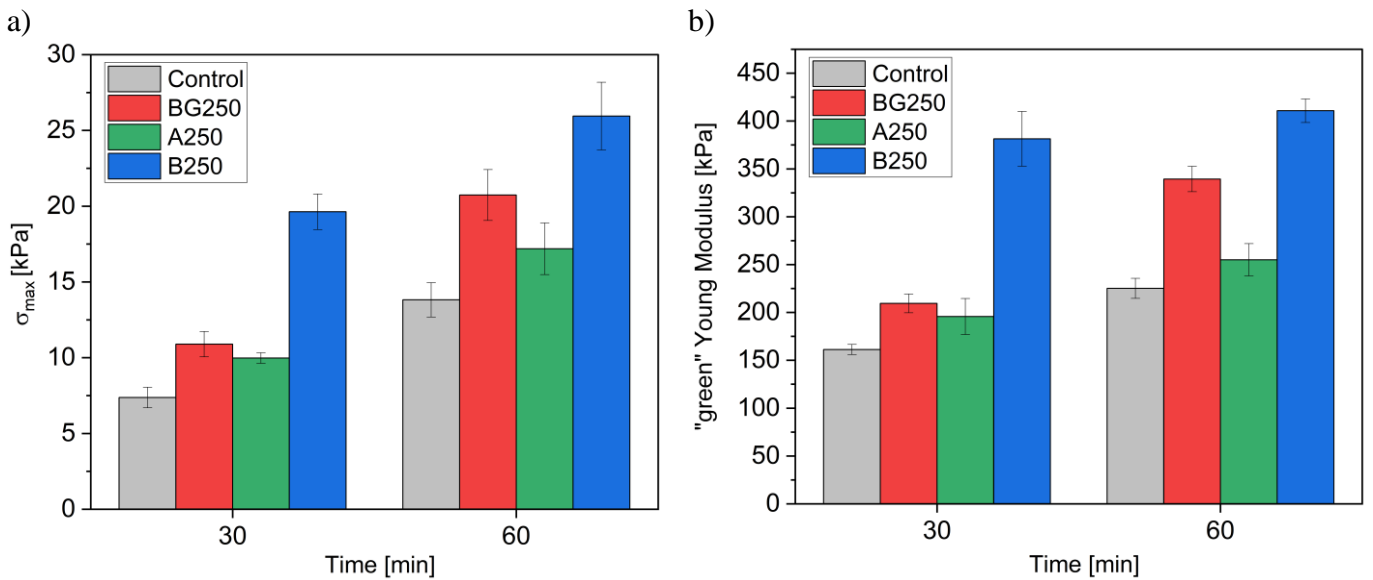

Figure 10. The comparison of green strength (a) and "green" Young's modulus (b) of 3DPC

## 4.6. Dynamic elastic modulus and early strength development

Results of dynamic elastic modulus ($E_{dyn}$) determination and development of early strength are presented in Figure 11. Differences between the onset of dynamic elastic modulus rise can be distinguished between specimens which can be associated with retardation of the hydration process as a result of introducing admixtures. Therefore, the development of dynamic elastic modulus is in line with the results of the calorimetric study (Table 3), confirming the case of specimens related to the reactivity of particles in the cementitious system. After 24 h, specimens BG250, A250, and B250 exhibited 20%, 13%, and 16% lower $E_{dyn}$ than the C (Figure 11a). The early strength values did not precisely align with the results obtained from $E_{dyn}$ and calorimetric tests. After the first 3 hours, a deficient compressive strength was recorded, likely due to the specimens still being in a plastic state and only beginning to set around that time (see Figure 11b). Starting from 6 hours, an apparent increase in the strength development of the Control specimen was noticed. Meanwhile, the second specimen to show early strength gain was B250. The modified specimens exhibited a delayed hydration process, remaining semi-plastic for up to 12 hours. Therefore, during the initial stages of specimen hardening, rheological performance plays a significant role in its ability to withstand stress. Based on the results from this and previous sections, incorporating BG and B particles in the printing process's initial stages positively impacts the fresh 3DPC and enhances its performance during the deposition stage. However, the material shows lower early strength gains after deposition, potentially reducing its ability to support layers.

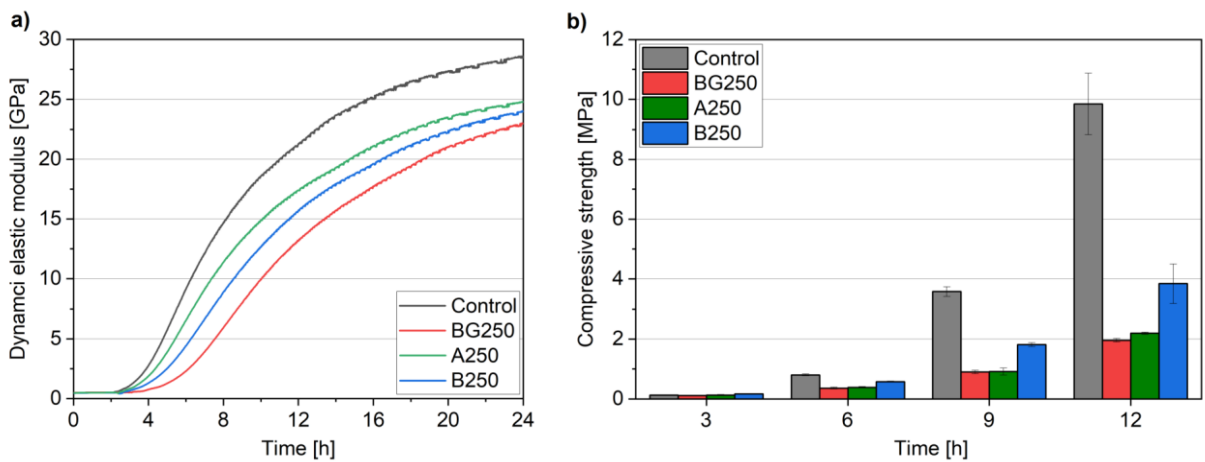

Figure 11. Dynamic elastic modulus (a) and early compressive strength (b) development of 3DPC

## 4.8. Failure pattern evaluation during flexural strength test and compressive strength performed on printed specimens

The experiment was focused on the value of horizontal deflection occurring in the sample during the flexural test, enabling the observation of the failure mechanism/pattern. The cracks developed from the lower layer subjected to tension and then extended across the entire cross-section of the sample. For each material group (A250, B250, BG250) and the control samples, five printed and three cast samples were made for each test, with the mean value taken as representative.

Figure 12 (a-b) depicts the comparison of flexural strength between cast and printed samples. It should be noted that cast specimens exhibited higher flexural strength values than printed ones, with differences ranging from 6.10% to 19.74%. The results for both sets of samples (cast and printed) indicate that the values for Control, A250, and B250 are almost identical. However, for BG250 specimens, the flexural strength is lower by 24.07% and 31.62% compared to the Control specimens for cast and printed samples, respectively.

The interlayer adhesion evaluation was performed using a splitting test for a more advanced analysis. The results are presented in Figure 12 (c-d). It should be noted that a similar trend to the flexural strength results is observed. For cast specimens, the Control sample exhibits the highest values of interlayer adhesion, and the difference between the A250 and B250 specimens is not statistically significant. The lowest values of interlayer adhesion were recorded for BG250 specimens, with values lower by 44.61% compared to Control samples. It should also be noted that the printing process influences the interlayer bonding, which is evident in all specimens. The splitting tensile force for printed specimens is lower than for cast specimens, ranging from 22.91% to 49.92%. The interlayer adhesion results for printed specimens range between 1.12 MPa (BG250) and 1.9 MPa (A250). Like cast specimens, the BG250 samples indicate the lowest values of interlayer adhesion; however, in this case, the difference compared to the Control samples is significantly smaller (23.69% lower than the Control samples).

Despite the differences in flexural strength and splitting tensile force (interlayer adhesion), the failure mechanism was identical for all tested samples. The failure occurred when the maximum force was reached during the test. One crack appeared, covering almost the entire cross-section of the sample simultaneously.

Figure 13 shows an example of the failure pattern analysis with a time interval of 1/100 s. The first photo shows no signs of damage, and the second photo shows one crack covering the entire cross-section. To sum up, despite the differences in flexural strength and splitting tensile force, the failure pattern is the same for all tested specimens. The phenomenon is explained that the specimens were printed without any time interval [34,35]. As reported above, a similar pattern was found for the compressive strength of 3D printed specimens. The printing process influenced the compressive strength values compared to the cast specimen, which is widely reported in the literature [43]. It includes particles with reduced compressive strength, similar for each sample, ranging between 29.20% and 35.56%. To sum up, the modification of the mix with pristine nano-sized $Bi_2O_3$ and $Gd_2O_3$ did not influence the failure pattern itself; however, it was confirmed that both types of $Bi_2O_3$/$Gd_2O_3$ coatings enable mitigate the retarded strength development when compared to 3DPC containing pristine $Bi_2O_3$/$Gd_2O_3$ particles.

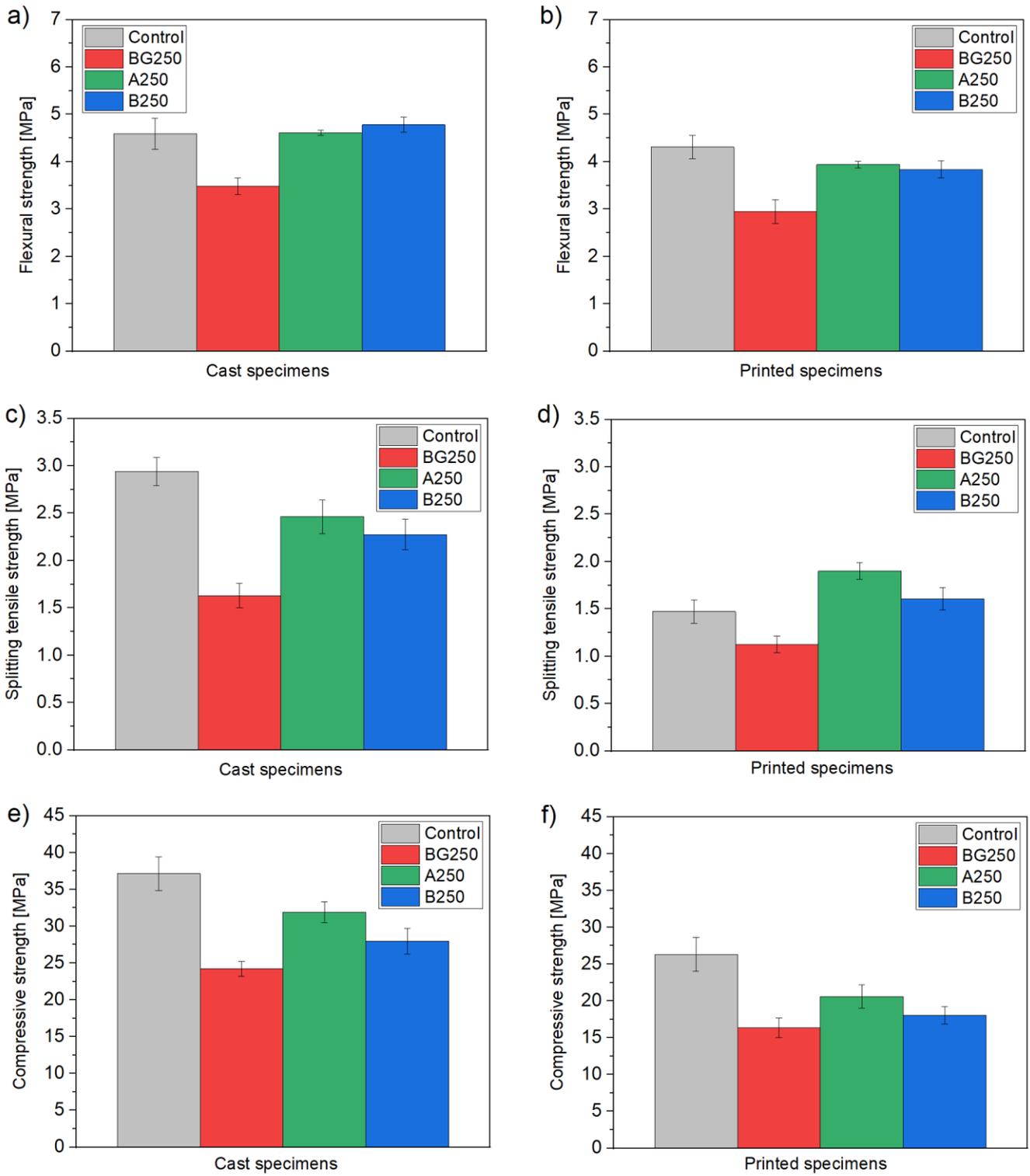

Figure 12. Comparison of flexural strength (a-b), splitting tensile strength (c-d), and compressive strength (e-f) of cast and printed specimens

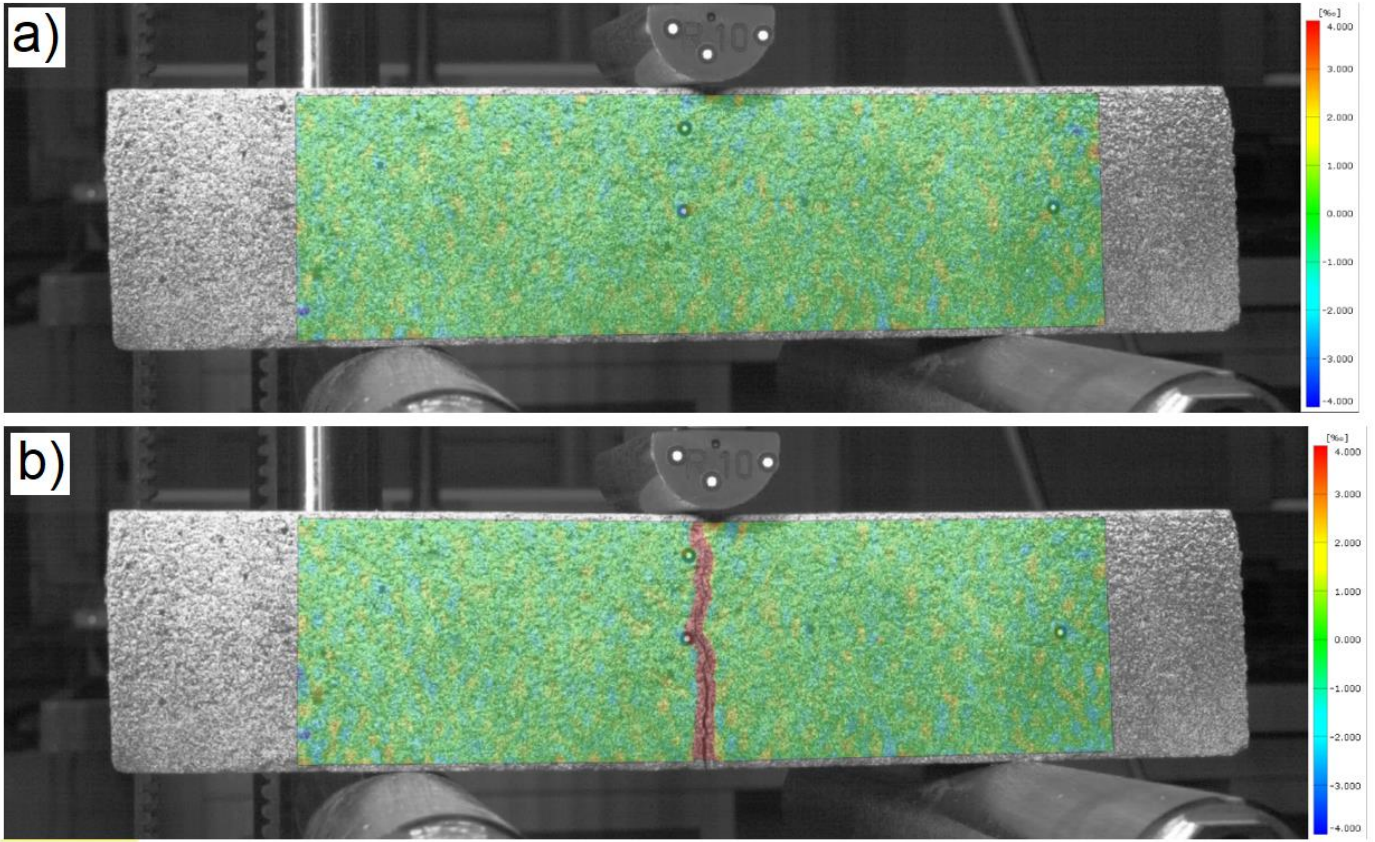

Figure 13. Failure pattern during flexural test: a) beginning of the test; b) end of the test

## 4.7. Microstructural characterization of 3D printed specimens (micro-CT and MIP)

Figure 14 shows the pore size distribution of printed C, BG250, A250, and B250 specimens. The measured porosity from these data is 12.70% (C), 13.70% (BG250), 12.09% (A250), and 9.90% (B250), and the result indicates that the B250 specimen exhibits relatively lower porosity than other cases. Although the BG250 shows the highest porosity, it contains more small pores than other cases, which could be attributed to the filling effect of singular particles of $Bi_2O_3$ and $Gd_2O_3$. Regarding pore size distribution, BG250 and A250 contain small pores, while B250 tends to have a relatively uniform size of the pores.

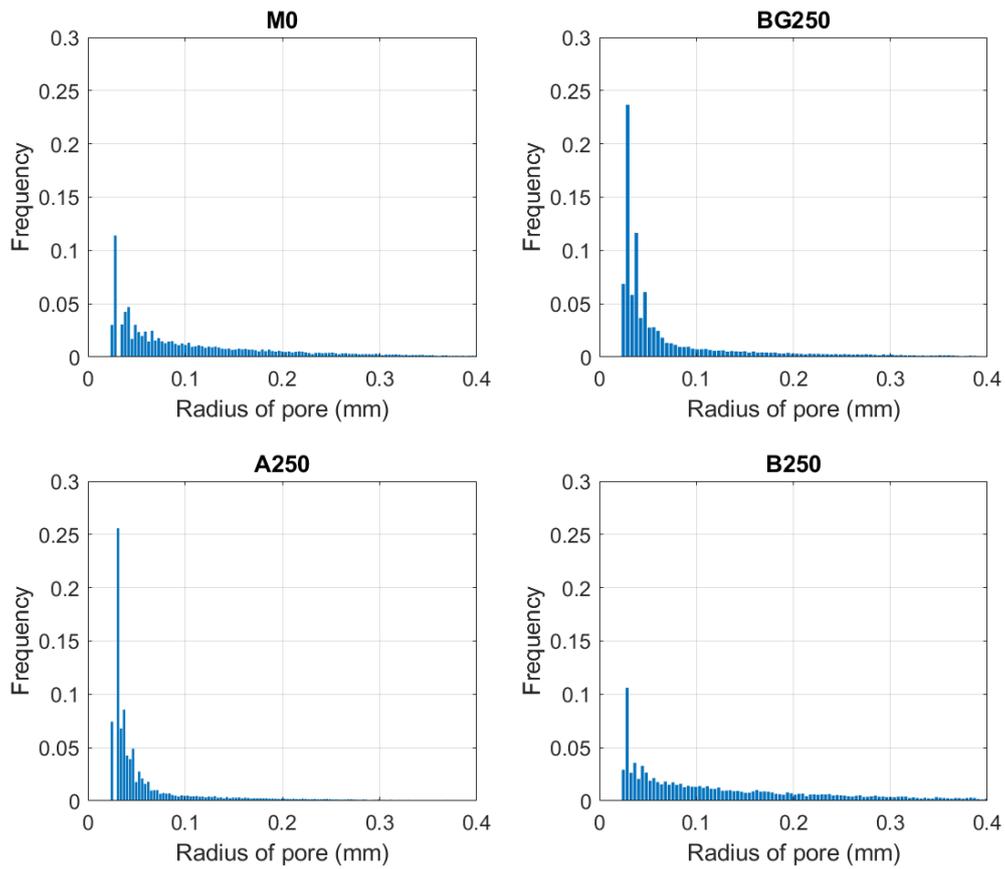

Figure 14. Pore size distribution of the specimens

To confirm the uniformity of the porosity, regional porosity according to the specimen height was examined, as shown in Figure 15. In this graph, it can be seen that all the materials are generally uniform in terms of pores, and there is no apparent intersectional defect in all specimens, which can be easily seen in 3D printed concrete; this indicates that the produced materials in this study are well designed without distinct heterogeneity, which can cause material defect or failure.

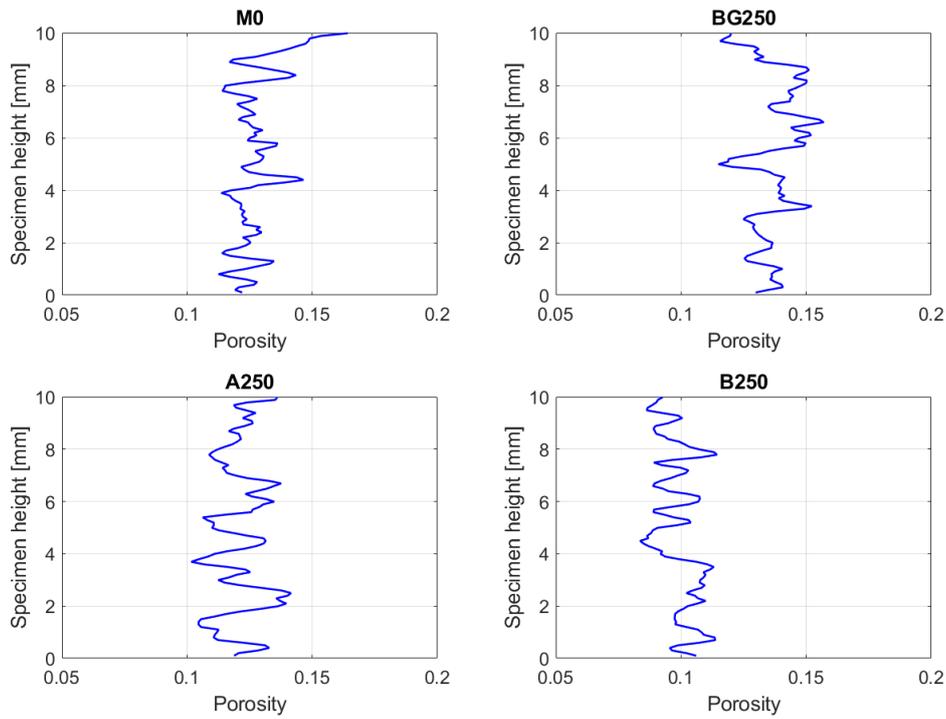

Figure 15. Porosity according to the specimen height

Concerning the interlayer effect in 3D printed concrete, in the previous works [12,33], it was confirmed that the pore shape is also affected by the material and printing process, which can significantly affect the material properties. To characterize it, a shape index, Wadell's sphericity, was adopted in this study. In Wadell's sphericity, the range is between 0 and 1, where 1 indicates a perfect sphere, while the value decreases as the anisotropy of an object increases. As shown in Figure 16, C shows the highest anisotropy, while B250 shows a relatively isotropic pore distribution compared to other cases, even in large pores. It presents that the B250 specimen contains more spherical pores that can reduce the directional discrepancy, particularly in the mechanical performance. Such results are in line with the buildability data presented in Section 4.5. confirming that the inclusion of silica-coated particles has a noticeable effect on the mixes' buildability, thus ensuring their pore structure is more uniform and spherical.

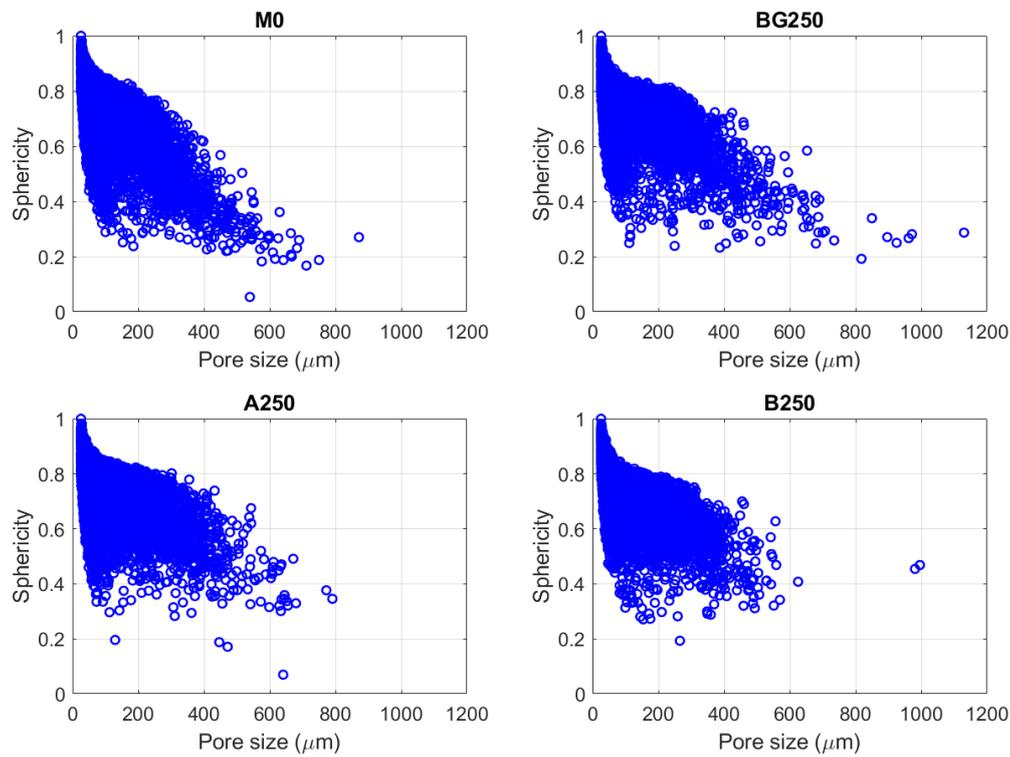

Figure 16. Wadell's sphericity of the specimens according to the pore sizes

The results of MIP tests are presented in Figure 17. Replacement of cement with particles resulted in a decrement in the average pore size and pore volume, represented by a slight shift of the pore peak between the control sample (C) and specimens containing admixtures. In the case of admixed specimens, it should be noted that the obtained materials have a more uniform pore structure than the C sample. For these materials, most pores range from 0.04 to 0.10 μm. Conversely, the control sample is dominated by pores with much larger diameters from 0.055 to 0.18 μm.

Comparing the microstructural properties obtained from MIP (Table 8), it should be noted that the use of ultra-fine structures resulted in a noticeable decrease in porosity in the tested measurement range. In the case of samples B250 and BG250, a decrease in total specific surface area was also noted. On the contrary, in the A250 sample, a decrease in porosity and an increase in specific surface area were noted compared to the Control sample. Therefore, it can be concluded that incorporating both uncoated and coated nanostructures benefits the MIP porosity compared to non-modified specimens.

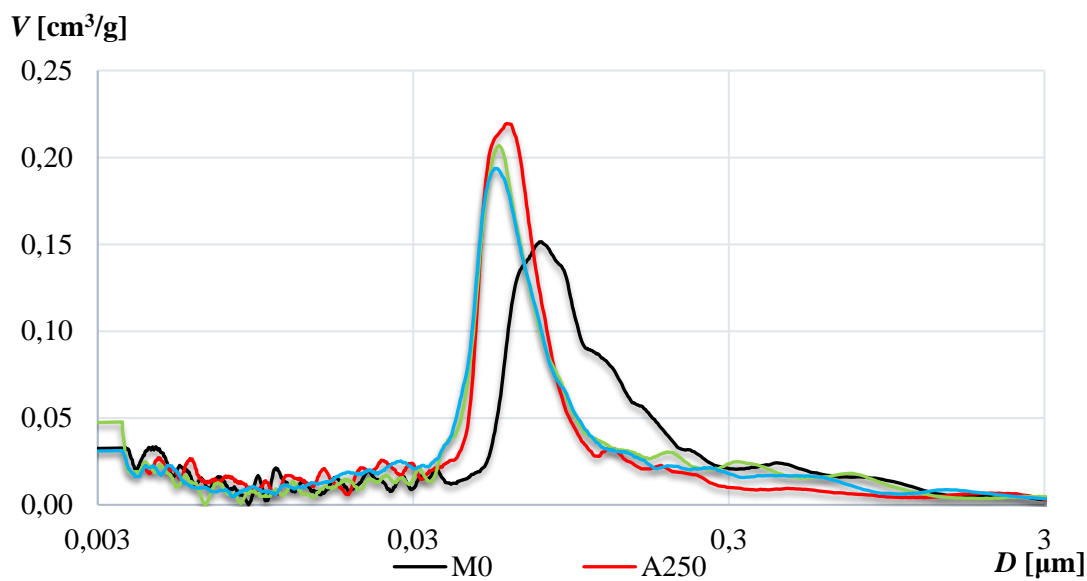

Figure 17. Log-differential pore size distribution (MIP)

Table 8. Porosity properties derived from mercury intrusion porosimetry

| Type | C | BG250 | A250 | B250 |
|---|---|---|---|---|
| Total specific surface [$m^2/g$] | 10.67 | 10.34 | 11.72 | 9.64 |
| Pore tortuosity [-] | 2.057 | 1.959 | 2.031 | 2.071 |
| Permeability [$10^{-4}$ $nm^2$] | 0.0007 | 0.0010 | 0.0005 | 0.0007 |
| Porosity (3 nm-300 μm) [%] | 19.93 | 18.70 | 17.35 | 19.49 |
| Average pore volume [$cm^3/g$] | 0.085 | 0.078 | 0.076 | 0.080 |
| for diameter [μm] | 0.038 | 0.036 | 0.031 | 0.039 |
| Median pore volume [$cm^3/g$] | 0.050 | 0.046 | 0.045 | 0.047 |
| for diameter [μm] | 0.098 | 0.065 | 0.063 | 0.069 |

### 4.8. Radiation shielding effectiveness
#### *4.8.1. Gamma-ray shielding effectiveness*

Figure 18a depicts the variation of the linear attenuation coefficient (LAC) with photon energy for different sample compositions. A clear trend emerges: as the incident photon energy increases, the ability of the materials to attenuate gamma rays (reflected by LAC) decreases. This well-established phenomenon is due to the changing interaction mechanisms between gamma rays and matter at different energy levels. Samples containing $Bi_2O_3$/$Gd_2O_3$ exhibit significantly higher γ-ray attenuation than the control sample (C). It aligns with expectations as the effective atomic number of the composite material increases with the presence of these elements with high atomic numbers. Interestingly, the LAC also seems affected by the specific metal coating method (A or B) employed on the samples. This observation aligns with our previous findings on the impact of coating methods [31]. Samples synthesized using method B exhibit superior attenuation compared to those synthesized with method A. However, both samples share the same composition (1:1 mass ratio $Bi_2O_3$/$Gd_2O_3$ particles coated with a silica shell *via* the sol-gel method) and contain comparable quantities of silica-coated, nano-sized $Bi_2O_3$ and $Gd_2O_3$ structures (Table 2), method B results in a significantly higher specific surface area (45.02 m²/g) compared to method A (8.01 m²/g). This larger surface area contributes to an increased cross-sectional area available for gamma-ray interaction, explaining the enhanced shielding effectiveness observed in samples prepared using method B. Additionally, the substantial increase in surface area (by about 562%) outweighs any potential drawbacks from the larger pore volume and minor decrease in density (around 2%) observed in B-type coated structures (Table 1).

Figure 17a reveals an intriguing observation: uncoated (pristine) $Bi_2O_3$/$Gd_2O_3$ nanoparticles (BG samples) exhibit higher gamma-ray attenuation compared to their silica-coated counterparts (A and B samples). While this might initially suggest a lack of effectiveness from the silica coating, a more likely explanation exists. As shown in Tables 1 and 2, the uncoated BG samples (7.94 g/cm³) contain higher concentrations of high-Z elements (36.7 kg/m³) compared to silica-coated samples (type A: 25.2 kg/m³, type B: 24.7 kg/m³). As an outcome, BG250 possessed higher density, followed by A250, B250, and C. This difference in composition and the higher density of BG samples leads to superior γ-ray attenuation.

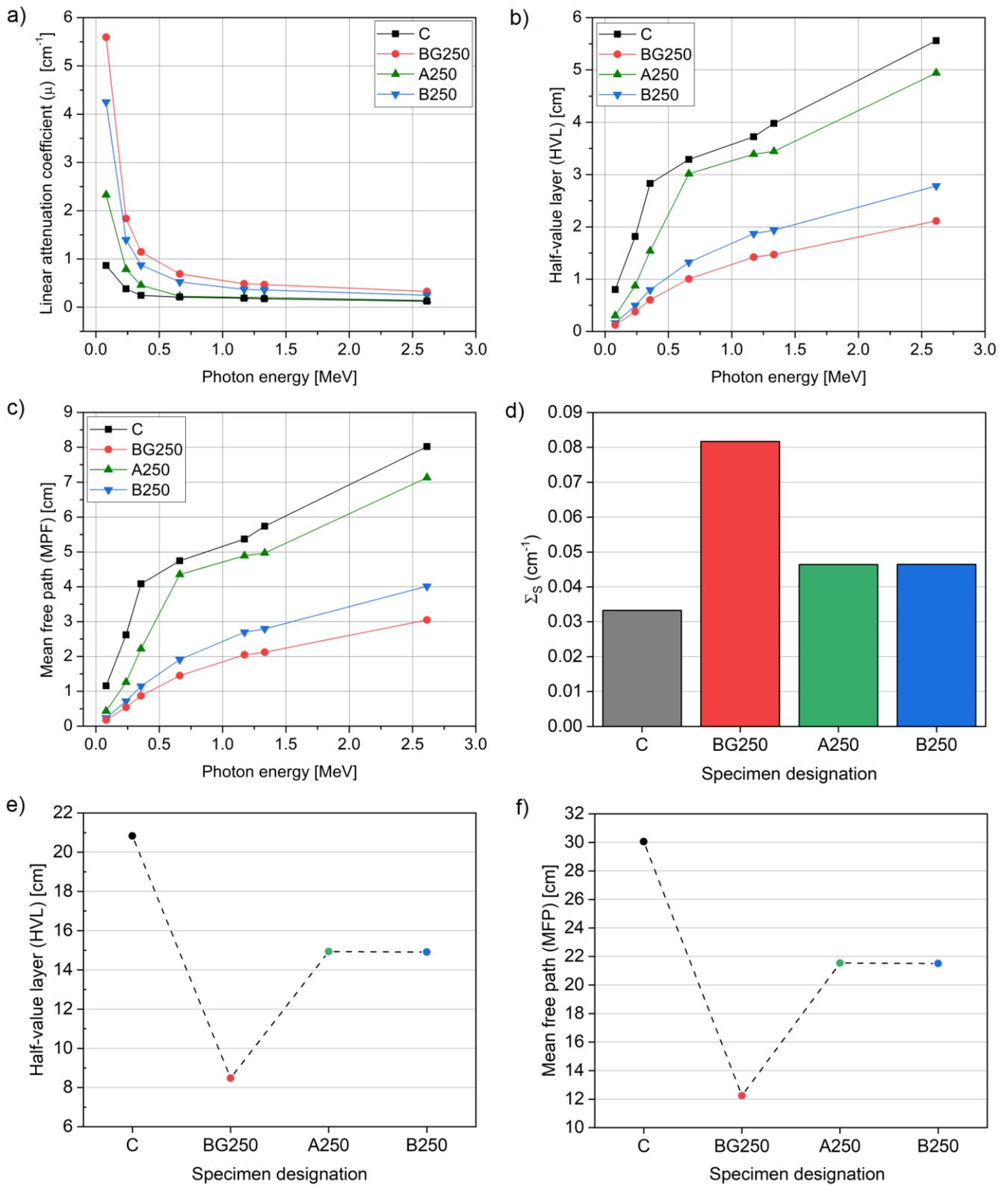

Figure 18. Linear attenuation coefficient (a), half-value layer (b), mean free path (c) determined for gamma-ray radiation and effective removal cross-section of fast neutron (d), half-value layer (e) and mean free path (f) determined for neutron radiation of 3DPC

As shown in Figure 18a, maintaining shielding effectiveness at higher energies requires increasing material thickness. This phenomenon is quantified by the half-value layer (HVL) as a function of photon energy (Figure 17b). Figure 17b illustrates the inverse relationship between HVL and shielding efficiency. 3DPC containing pristine particles and those synthesized using method B exhibit a lower HVL, confirming their ability to provide adequate shielding with a reduced material thickness. Notably, Figure 18b reveals that the HVL values of the A250 sample are considerably lower than those of the B250 sample, and this disparity becomes more pronounced for higher energy levels. Figure 18c illustrates the same concept by plotting the Mean Free Path (MFP), defined as the average distance travelled between successive collisions (MFP = $1/\mu$). The recorded MFP values were 1.16 cm, 0.18 cm, 0.24 cm, and 0.44 cm for C, BG250, B250, and A250, respectively.

### 4.8.2. *Slow neutron shielding effectiveness*

Figure 18d depicts the total neutron slow cross-section values ($\Sigma_S$, cm$^{-1}$) of tested 3DPC mixes. Similar to gamma-ray studies, adding pristine and silica-coated $Bi_2O_3/Gd_2O_3$ nanoparticles to the 3DCP specimens improved neutron attenuation compared with the control sample (C). This improvement can be attributed to two factors. First, $Gd_2O_3$ has a high capture cross-section for slow and thermal neutrons. Second, adding $Bi_2O_3$ increases the material density, further enhancing neutron interactions. Consistent with the γ-ray findings, structure B with silica coating showed a more remarkable improvement in neutron shielding capability than structure A. As expected, sample BG250 exhibits the highest overall $\Sigma_S$ (improvement 2.46 times higher when compared to C). In contrast, samples A250 and B250 have comparable $\Sigma_S$ values. This is because the relatively higher $Bi_2O_3/Gd_2O_3$ content of A250 (25.2 kg/m$^3$ compared to 24.7 kg/m$^3$ for B250, as shown in Table 2 in the manuscript ) compensates for the improvement introduced by coating method B. Half-Value Layer and Mean Free Path exhibit an inverse relationship with shielding effectiveness. Figures 17e and 17f illustrate lower HVL and MFP values for the BG sample. It translates to requiring less BG sample material to achieve the same level of protection. The order of HVL and MFP from highest to lowest is C > A > B > BG. This finding demonstrates the positive influence of adding $Bi_2O_3/Gd_2O_3$ nanoparticles and the superiority of coating method B compared to method A.

**CONCLUSIONS**

The obtained results demonstrated several aspects of the use of new types of silica-coated $Bi_2O_3$-$Gd_2O_3$ structures for 3DPC applications, and the following conclusions can be drawn:

- Isothermal calorimetry studies confirmed that introducing pristine bismuth oxide-gadolinium oxide (BG) particles dramatically decreased the peak magnitude and occurrence. The introduction of A or B-type silica-coating on the surface of particles enabled to overcome the retarding effect on hydration attributed to $Bi_2O_3$, and thus the occurrence of peaks was only slightly delayed when compared to the Control specimen. At the same time, cumulative heat values after seven days remained similar to that of the C specimen as well. From a hydration perspective, coating type A was more efficient than B. As an effect of the delayed hydration process, specimens containing non-coated BG particles exhibited lower strength, especially in the first two days of hydration, while after seven days, the effect was negligible. The introduction of silica coating enabled this effect to be overcome. After 24 hours, the best-performing mixes from each group (BG250, A250, and B250) exhibited 35%, 14%, and 26% lower compressive strength when compared to C.
- The incorporation of plain and coated BG additives leads to a substantial increase in the static and dynamic yield shear stresses of the mixtures. The most favorable amount of admixture was 2.5 vol% in all cases. Coating method B results in a significantly higher specific surface area (45.02 m²/g) than method A (8.01 m²/g). Fine silica particle size and larger surface area contributed to the most favorable effect on the fresh 3DPC and enhanced the performance of 3DPC during the deposition state when BG particles coated with method B were used (B250). The results indicated that adding BG particles, both with and without coating, can substantially improve the thixotropy of composites compared to that of the control sample. This is attributed to the flocculation induced by the BG particles due to their nano-sized nature, resulting in a higher thixotropic loop area. As per the results, it can be inferred that the mixtures containing 2.5% BG particles (i.e., BG250, A250, and B250) exhibited the highest levels of thixotropy, regardless of the coating type.
- The presented results indicated that incorporating pristine and silica-coated $Bi_2O_3/Gd_2O_3$ structures benefits the "green" mechanical properties and buildability of 3D printed mixes. The highest values of green Young's modulus and green strength were observed in the $Bi_2O_3/Gd_2O_3$ coated using method B (B250). For mixes tested after 30 min, the green strength and green Young's modulus increased by up to 62.4% and 57.7%, respectively, compared to the Control sample. For mixes tested after 60 min, these increases were up to 46.7% and 45.2%, respectively.

- However, after the deposition state, the 3DPC incorporated with non-coated and coated particles exhibited lower early strength gains and, thus, potentially lower ability to withstand layers. Introducing silica coating reduced the negative impact of BG particles and faster setting time and strength gaining.
- Micro-CT investigations confirmed that silica-coated structures substantially affect the alteration of pore structures (especially B-type coating), ensuring the lowest pore anisotropy distribution (more spherical pores) and lower porosity. At the same time, MIP confirmed the structure's refinement; thus, lower porosity and median pore diameter were reported in specimens containing uncoated and coated BG particles.
- 2.5 vol.% replacement of cement with structures (non-coated and coated) was found to be most beneficial for the overall performance (printability and mechanical performance) of 3DPC and enables the introduction of the structures without altering the mixture design. High replacement levels (5 vol%) of cement with particles resulted in noticeable material agglomeration (confirmed by SEM/EDS), which in turn resulted in decrement of the rheological performance, the consistency as well as reduction of open time (extrudability). The magnitude of these phenomena was increased when coated specimens were used.
- Modifying the mix with pristine or silica-coated $Bi_2O_3/Gd_2O_3$ structures did not affect the observed failure pattern using DIC after one day of curing. However, distinctively lower mechanical performance of 3D printed specimens was found in specimens containing pristine $Bi_2O_3/Gd_2O_3$ structures. Including coated specimens enables the mitigated low strength gain in 3D printed samples.
- Due to the highest concentrations of high-Z elements, mixes with BG particles represented the best radiation shielding performance (gamma-ray and neutron). The introduction of uncoated and coated BG structures resulted in an increment of the 3DPC density. Samples synthesized using method B exhibit superior attenuation compared to those synthesized with method A. The higher surface area of specimens coated with method B contributes to an increased cross-sectional area available for gamma-ray interaction, explaining the enhanced shielding effectiveness. The order of HVL and MFP from highest to lowest is C > A > B > BG. This finding demonstrates the positive influence of adding $Bi_2O_3/Gd_2O_3$ particles to the mixtures.

**Funding:** This research was funded in whole by the National Science Centre, Poland within Project no. 2020/39/D/ST8/00975 (SONATA-16).

# References

[1] K. Cuevas, J. Strzałkowski, J.-S. Kim, C. Ehm, T. Glotz, M. Chougan, S.H. Ghaffar, D. Stephan, P. Sikora, Towards development of sustainable lightweight 3D printed wall building envelopes – Experimental and numerical studies, Case Studies in Construction Materials 18 (2023) e01945. https://doi.org/10.1016/j.cscm.2023.e01945.

[2] G. Davis, C. Montes, S. Eklund, Preparation of lunar regolith based geopolymer cement under heat and vacuum, Advances in Space Research 59 (2017) 1872–1885. https://doi.org/10.1016/j.asr.2017.01.024.

[3] D. Juračka, J. Katzer, J. Kobaka, I. Świca, K. Seweryn, Concept of a 3D-Printed Voronoi Egg-Shaped Habitat for Permanent Lunar Outpost, Applied Sciences 13 (2023) 1153. https://doi.org/10.3390/app13021153.

[4] S. Ma, Y. Jiang, S. Fu, P. He, C. Sun, X. Duan, D. Jia, P. Colombo, Y. Zhou, 3D-printed Lunar regolith simulant-based geopolymer composites with bio-inspired sandwich architectures, Journal of Advanced Ceramics 12 (2023) 510–525. https://doi.org/10.26599/JAC.2023.9220700.

[5] S.C. Ligon, R. Liska, J. Stampfl, M. Gurr, R. Mülhaupt, Polymers for 3D Printing and Customized Additive Manufacturing, Chem. Rev. 117 (2017) 10212–10290. https://doi.org/10.1021/acs.chemrev.7b00074.

[6] Y. Cao, H. Yang, K. Wan, D. Li, Q. He, H. Wu, High-performance PEEK composite materials research on 3D printing for neutron and photon radiation shielding, Composites Part A: Applied Science and Manufacturing 185 (2024) 108352. https://doi.org/10.1016/j.compositesa.2024.108352.

[7] B. Gultekin, F. Bulut, H. Yildiz, H. Us, H. Ogul, Production and investigation of 3D printer ABS filaments filled with some rare-earth elements for gamma-ray shielding, Nuclear Engineering and Technology 55 (2023) 4664–4670. https://doi.org/10.1016/j.net.2023.09.009.

[8] M. Elsafi, M.A. El-Nahal, M.I. Sayyed, I.H. Saleh, M.I. Abbas, Novel 3-D printed radiation shielding materials embedded with bulk and nanoparticles of bismuth, Sci. Rep. 12 (2022) 12467. https://doi.org/10.1038/s41598-022-16317-w.

[9] S.J. Talley, T. Robison, A.M. Long, S.Y. Lee, Z. Brounstein, K.-S. Lee, D. Geller, E. Lum, A. Labouriau, Flexible 3D printed silicones for gamma and neutron radiation shielding, Radiation Physics and Chemistry 188 (2021) 109616. https://doi.org/10.1016/j.radphyschem.2021.109616.

[10] B. Kanagaraj, N. Anand, A. Diana Andrushia, M.Z. Naser, Recent developments of radiation shielding concrete in nuclear and radioactive waste storage facilities – A state of the art review, Construction and Building Materials 404 (2023) 133260. https://doi.org/10.1016/j.conbuildmat.2023.133260.

[11] G. Tyagi, A. Singhal, S. Routroy, D. Bhunia, M. Lahoti, Radiation Shielding Concrete with alternate constituents: An approach to address multiple hazards, J. Hazard. Mater. 404 (2021) 124201. https://doi.org/10.1016/j.jhazmat.2020.124201.

[12] K. Federowicz, M. Techman, S. Skibicki, M. Chougan, A.M. El-Khayatt, H.A. Saudi, J. Błyszko, M. Abd Elrahman, S.-Y. Chung, P. Sikora, Development of 3D printed heavyweight concrete (3DPHWC) containing magnetite aggregate, Materials & Design 233 (2023) 112246. https://doi.org/10.1016/j.matdes.2023.112246.

[13] P. Gokul, J. Ashok Kumar, R. Preetha, S. Chattopadhyaya, K.M. Mini, Additives in concrete to enhance neutron attenuation characteristics – A critical review, Results in Engineering 19 (2023) 101281. https://doi.org/10.1016/j.rineng.2023.101281.

[14] B.K. Soni, R. Makwana, S. Mukherjee, S.S. Barala, S. Parashari, R. Chauhan, A.S. Jodha, K. Katovsky, Novel concrete compositions for γ-rays and neutron shielding using WC and $B_4C$, Results in Materials 10 (2021) 100177. https://doi.org/10.1016/j.rinma.2021.100177.

[15] I.M. Nikbin, M. Shad, G.A. Jafarzadeh, S. Dezhampanah, An experimental investigation on combined effects of nano-$WO_3$ and nano-$Bi_2O_3$ on the radiation shielding properties of magnetite concretes, Progress in Nuclear Energy 117 (2019) 103103. https://doi.org/10.1016/j.pnucene.2019.103103.

[16] O. Bawazeer, K. Makkawi, Z.B. Aga, H. Albakri, N. Assiri, K. Althagafy, A.-W. Ajlouni, A review on using nanocomposites as shielding materials against ionizing radiation, J.Umm Al-Qura Univ. Appll. Sci. 9 (2023) 325–340. https://doi.org/10.1007/s43994-023-00042-9.

[17] H.M.H. Zakaly, G. ALMisned, S.A.M. Issa, V. Ivanov, H.O. Tekin, Towards a better understanding of filler size on radiation shielding enhancement: impact of micro- and nano-$WO_3$/PbO particle reinforcement on ILC concrete, J Aust Ceram Soc 59 (2023) 127–135. https://doi.org/10.1007/s41779-022-00818-y.

[18] M.A. El-Nahal, M. Elsafi, M.I. Sayyed, M.U. Khandaker, H. Osman, B.H. Elesawy, I.H. Saleh, M.I. Abbas, Understanding the Effect of Introducing Micro- and Nanoparticle Bismuth Oxide (Bi2O3) on the Gamma Ray Shielding Performance of Novel Concrete, Materials (Basel) 14 (2021). https://doi.org/10.3390/ma14216487.

[19] M.M. Gouda, A.M. El-Khatib, M.I. Abbas, S.M. Al-Balawi, M.T. Alabsy, Gamma Attenuation Features of White Cement Mortars Reinforced by Micro/Nano Bi2O3 Particles, Materials (Basel) 16 (2023). https://doi.org/10.3390/ma16041580.

[20] Y. Xia, T. Yang, G. Pan, S. Xu, L. Yao, Preparation of Core–Shell Structure W/Gd2O3 and Study on the Properties of Radiation Protection Materials, Coatings 12 (2022) 851. https://doi.org/10.3390/coatings12060851.

[21] P. Zhang, C. Jia, J. Li, W. Wang, Shielding composites for neutron and gamma-radiation with Gd2O3@W core-shell structured particles, Materials Letters 276 (2020) 128082. https://doi.org/10.1016/j.matlet.2020.128082.

[22] I.G. Alhindawy, M.I. Sayyed, A.H. Almuqrin, K.A. Mahmoud, Optimizing gamma radiation shielding with cobalt-titania hybrid nanomaterials, Sci. Rep. 13 (2023) 8936. https://doi.org/10.1038/s41598-023-33864-y.

[23] S.M. Ibrahim, M. Heikal, N.R. Abdelwahab, O.A. Mohamed, Fabricated CeO2/ZrO2 nanocomposite to improve thermal resistance, mechanical characteristics, microstructure and gamma radiation shielding of OPC composite cement pastes, Construction and Building Materials 392 (2023) 131971. https://doi.org/10.1016/j.conbuildmat.2023.131971.

[24] Q. Li, N.J. Coleman, Impact of Bi2O3 and ZrO2 Radiopacifiers on the Early Hydration and C-S-H Gel Structure of White Portland Cement, J. Funct. Biomater. 10 (2019). https://doi.org/10.3390/jfb10040046.

[25] K. Cendrowski, K. Federowicz, M. Techman, M. Chougan, T. Kędzierski, M. Sanytsky, E. Mijowska, P. Sikora, Enhancing the Fresh and Early Age Performances of Portland Cement Pastes via Sol-Gel Silica Coating of Metal Oxides (Bi2O3 and Gd2O3), Coatings 13 (2023) 1698. https://doi.org/10.3390/coatings13101698.

[26] S.L. Ho, H. Yue, T. Tegafaw, M.Y. Ahmad, S. Liu, S.-W. Nam, Y. Chang, G.H. Lee, Gadolinium Neutron Capture Therapy (GdNCT) Agents from Molecular to Nano: Current Status and Perspectives, ACS Omega 7 (2022) 2533–2553. https://doi.org/10.1021/acsomega.1c06603.

[27] M. Bolhassani, M. Sayyahmanesh, A study on mechanical properties of cement paste using magnetite-silica nano-composites, Advances in Cement Research 27 (2015) 571–580. https://doi.org/10.1680/adcr.14.00106.

[28] B. Han, Z. Li, L. Zhang, S. Zeng, X. Yu, B. Han, J. Ou, Reactive powder concrete reinforced with nano SiO2-coated TiO2, Construction and Building Materials 148 (2017) 104–112. https://doi.org/10.1016/j.conbuildmat.2017.05.065.

[29] P. Sikora, M. Abd Elrahman, S.-Y. Chung, K. Cendrowski, E. Mijowska, D. Stephan, Mechanical and microstructural properties of cement pastes containing carbon nanotubes and carbon nanotube-silica core-shell structures, exposed to elevated temperature, Cement and Concrete Composites 95 (2019) 193–204. https://doi.org/10.1016/j.cemconcomp.2018.11.006.

[30] Z. Pi, H. Xiao, R. Liu, H. Li, Combination usage of nano-SiO2-coated steel fiber and silica fume and its improvement effect on SFRCC, Composites Part B: Engineering 221 (2021) 109022. https://doi.org/10.1016/j.compositesb.2021.109022.

[31] K. Cendrowski, K. Federowicz, M. Techman, M. Chougan, A.M. El-Khayatt, H.A. Saudi, T. Kędzierski, E. Mijowska, J. Strzałkowski, D. Sibera, M. Abd Elrahman, P. Sikora, Functional Bi2O3/Gd2O3 Silica-Coated Structures for Improvement of Early Age and Radiation Shielding Performance of Cement Pastes, Nanomaterials (Basel) 14 (2024). https://doi.org/10.3390/nano14020168.

[32] K. Cendrowski, P. Sikora, B. Zielinska, E. Horszczaruk, E. Mijowska, Chemical and thermal stability of core-shelled magnetite nanoparticles and solid silica, Applied Surface Science 407 (2017) 391–397. https://doi.org/10.1016/j.apsusc.2017.02.118.

[33] Y.A. Al-Noaimat, M. Chougan, A. Albar, S. Skibicki, K. Federowicz, M. Hoffman, D. Sibera, K. Cendrowski, M. Techman, J.N. Pacheco, S.-Y. Chung, P. Sikora, M. Al-Kheetan, S.H. Ghaffar, Recycled brick aggregates in one-part alkali-activated materials: Impact on 3D printing performance and material properties, Developments in the Built Environment 16 (2023) 100248. https://doi.org/10.1016/j.dibe.2023.100248.

[34] S. Skibicki, P. Szewczyk, J. Majewska, D. Sibera, E. Ekiert, S.-Y. Chung, P. Sikora, The effect of interlayer adhesion on stress distribution in 3D printed beam elements, Journal of Building Engineering 87 (2024) 109093. https://doi.org/10.1016/j.jobe.2024.109093.

[35] R. Wolfs, F.P. Bos, T. Salet, Hardened properties of 3D printed concrete: The influence of process parameters on interlayer adhesion, Cement and Concrete Research 119 (2019) 132–140. https://doi.org/10.1016/j.cemconres.2019.02.017.

[36] P. Sikora, A.M. El-Khayatt, H.A. Saudi, S.-Y. Chung, D. Stephan, M. Abd Elrahman, Evaluation of the effects of bismuth oxide (Bi2O3) micro and nanoparticles on the mechanical, microstructural and γ-ray/neutron shielding properties of Portland cement pastes, Construction and Building Materials 284 (2021) 122758. https://doi.org/10.1016/j.conbuildmat.2021.122758.

[37] T. Piotrowski, J. Glinicka, M.A. Glinicki, P. Prochoń, Influence of gadolinium oxide and ulexite on cement hydration and technical properties of mortars for neutron radiation shielding purposes, Construction and Building Materials 195 (2019) 583–589. https://doi.org/10.1016/j.conbuildmat.2018.11.076.

[38] M. Azeem, M.A. Saleem, A Raman Spectroscopic Study of Calcium Silicate Hydrate (CSH) in the Cement Matrix with CNTs and Oxide Additives, Journal of Spectroscopy 2022 (2022) 1–6. https://doi.org/10.1155/2022/2281477.

[39] Y.W.D. Tay, Y. Qian, M.J. Tan, Printability region for 3D concrete printing using slump and slump flow test, Composites Part B: Engineering 174 (2019) 106968. https://doi.org/10.1016/j.compositesb.2019.106968.

[40] R. Jayathilakage, P. Rajeev, J. Sanjayan, Rheometry for Concrete 3D Printing: A Review and an Experimental Comparison, Buildings 12 (2022) 1190. https://doi.org/10.3390/buildings12081190.

[41] M. Chen, L. Yang, Y. Zheng, Y. Huang, L. Li, P. Zhao, S. Wang, L. Lu, X. Cheng, Yield stress and thixotropy control of 3D-printed calcium sulfoaluminate cement composites with metakaolin related to structural build-up, Construction and Building Materials 252 (2020) 119090. https://doi.org/10.1016/j.conbuildmat.2020.119090.

[42] Z. Zhao, T. Qi, W. Zhou, D. Hui, C. Xiao, J. Qi, Z. Zheng, Z. Zhao, A review on the properties, reinforcing effects, and commercialization of nanomaterials for cement-based materials, Nanotechnology Reviews 9 (2020) 303–322. https://doi.org/10.1515/ntrev-2020-0023.

[43] A. Baktheer, M. Classen, A review of recent trends and challenges in numerical modeling of the anisotropic behavior of hardened 3D printed concrete, Additive Manufacturing 89 (2024) 104309. https://doi.org/10.1016/j.addma.2024.104309.